\documentclass[12pt]{iopart}
\usepackage{iopams}  
\usepackage{graphicx}

\usepackage[normalem]{ulem}\usepackage{soul}

\usepackage{color}

\eqnobysec

\newcommand{\PT}{$\mathcal{PT}$}
\newcommand{\p}{$\mathcal{P} \; $}
\newcommand{\T}{$\mathcal{T} \;$}

\newcommand{\be}{\begin{equation}}
\newcommand{\ee}{\end{equation}}

\newcommand{\sech}{\mathrm{sech}}

\begin{document}

\title{Spinor solitons and their 
 $\mathcal{PT}$-symmetric offspring}

\author{N. V. Alexeeva$^{1,2,3}$
I. V.  Barashenkov$^{1,2,3}$  and A.  Saxena$^{2}$}

\address
{$^1$ Department of Mathematics, University of Cape Town, Private Bag X3,
 Rondebosch 7701, South Africa \\
$^2$ Center for Nonlinear Studies, Los Alamos National Laboratory,  MSB258
Los Alamos, New Mexico 87545, USA \\
 $^3$ Department of Physics, University of Bath, Claverton Down, 
Bath BA2 7AY, UK}

\ead{Nora.Alexeeva@uct.ac.za,
Igor.Barashenkov@uct.ac.za,  avadh@lanl.gov}
\vspace{10pt}

\begin{abstract}

Although the  spinor field in (1+1) dimensions  has the right structure to model a dispersive bimodal 
 system with gain and loss, the plain addition of gain to one component of the field and
loss to the other one results in an unstable dispersion relation. 
In this paper, we advocate a different  recipe for the $\mathcal{PT}$-symmetric extension of spinor models
--- the recipe that does not produce instability of the linear Dirac equation. 
 Having exemplified the physical origins of the $\mathcal P$- and $\mathcal T$-breaking terms, we 
consider the extensions of three U(1)-invariant spinor models with cubic nonlinearity.
Of these,
the \PT-symmetric extension of the Thirring model is shown to be completely integrable
and possess infinitely many conserved quantities.
The  \PT-symmetric Gross-Neveu equation conserves energy and momentum but does not conserve charge. 
The third model is introduced  for the purpose of comparison with the previous two;
its \PT-symmetric extension  has no conservation laws at all. 
 Despite this  dramatic difference in the integrability and conservation properties, 
  all three \PT-symmetric models are
 shown to have exact soliton solutions. Similar to the solitons of the extended Thirring and Gross-Neveu equations,  the solitons of the new model 
 are found to be stable --- except for a narrow band of frequencies adjacent to the soliton existence boundary.
The persistence under the $\mathcal P$- and $\mathcal T$-breaking perturbations as well as the  prevalence of stability
highlight a remarkable sturdiness of spinor solitons in (1+1) dimensions.

\end{abstract}

%
%
%
%
%

\section{Introduction} 

Rooted in  the nonhermitian quantum mechanics \cite{Bender},
 the \PT-symmetric extensions of conservative systems are becoming increasingly relevant in applied disciplines
 \cite{PT_Focus}.
 In optics, the \PT-symmetric structures are brought about by the balanced application of gain and loss.
Behaviours afforded by the \PT-symmetric arrangements and 
  unattainable in standard set-ups, include
 the
unconventional beam refraction \cite{refraction,Regensburger}, 
 loss-induced  transparency \cite{Guo}, and
nonreciprocal light propagation \cite{Nonreciprocal_propagation}.   
\PT-symmetric systems are expected to promote an
   efficient control of light, including
all-optical low-threshold  switching \cite{RKEC,switch}                     
and unidirectional
invisibility \cite{RKEC,
Regensburger,invisibility}.
 There is also  a  growing interest   in the context of plasmonics \cite{plasmonics},   optomechanical systems \cite{OM}  and metamaterials \cite{Lazarides}.

Recent theoretical analyses of the distributed \PT-symmetric systems  were exploiting various forms of the nonlinear 
Schr\"odinger equation,  continuous \cite{NLS_single,NLS_dimer} or discrete \cite{NLS_disc}. 
The fundamental object under scrutiny was a particle-like bunch of energy --- a soliton, breather or localised internal mode \cite{reviews}.
The present study is concerned with  a \PT-symmetric extension of 
 another workhorse of the   wave theory, namely 
the nonlinear Dirac equation.

A particularly simple \PT-symmetric Schr\"odinger system  consists of two coupled modes, 
of which one component gains  and the other one loses energy  at an equal rate \cite{NLS_dimer}. 
Like this  Schr\"odinger dimer, 
the (1+1)-dimensional Dirac field consists of  two symmetrically arranged components and
 is ideally suited for the symmetric application of
gain and loss. The Lorentz invariance of the Dirac field  is an additional built-in symmetry
which can be preserved by the gain and loss terms. The two components of the field transform as 
a Lorentz spinor.

As the authors of  the Schr\"odinger-based studies, we will be focussing on the 
 localised solutions of the nonlinear Dirac  equation. That is to say, our interest lies in the \PT-symmetric spinor solitons.

The  nonlinear Dirac equation  with the scalar self-interaction
was introduced
 by Ivanenko  \cite{R1};
 the vector self-interaction is due to Thirring  \cite{R3}. In the 1950s,
 Ivanenko \cite{R6} and, independently, Heisenberg \cite{R2} adopted the former equation as a basis for the unified nonlinear field theory.
 In the 1970s,
  Soler tried to use the  scalar self-interaction model to describe extended nucleons 
 \cite{R4} while Gross and Neveu employed  its one-dimensional version  to explain the quark confinement \cite{R5}.
Outside  the realm of elementary particles, the Gross-Neveu theory was utilised in the study of polymers \cite{BC}
 while the massive Thirring model appeared in the context of optical gratings \cite{gratings}. 
 Mathematically, the Thirring model was shown to be completely integrable via the Inverse Scattering Transform \cite{Mikhailov}.

 The latest wave of interest in the Dirac equation in the condensed-matter context  concerns
 the electronic structure of two-dimensional 
materials
   graphene and silicene  \cite{R8}
   as well as the transition metal dichalcogenides \cite{R9}.   
 A closely related topic is  bosonic evolution in honeycomb 
lattices \cite{R7}.
The recent applications of the Dirac equation in optics, are to the light propagation in honeycomb photorefractive lattices 
  (photonic graphene) \cite{R10} and conical diffraction in such  structures \cite{R11}.  
  The spin-orbit coupled Bose-Einstein condensates is yet another area of utilisation of 
  the Dirac-type equations  \cite{R12}. 
We should also mention  a renewed interest in the stability properties of the Dirac solitons \cite{new_stability,Berkolaiko,Num_GN,Dima,Taras}
  that have been commonly seen as a mystery \cite{old_stability} despite some early progress \cite{BPZ}.

There can be a variety of physically meaningful 
 \PT-symmetric perturbations of the  Dirac equation --- that is,
 perturbations by 
  terms
that   break each of 
the parity- and time-reversal symmetries of the equation but 
remain invariant under the joint action of the \p
and \T operators. 
The present study is confined  to \PT-symmetric perturbations
that preserve the invariance under the Lorentz rotations 
(velocity boosts)
and the U(1) gauge transformations.

We scrutinise the general recipe of the 
\PT-symmetric extension of the Dirac equation,
explore properties of the extended nonlinear models,
 construct   exact soliton solutions
 and examine their stability. 
To
 crystallise  common properties of the \PT-symmetric spinor solitons, 
we consider three different spinor equations together with their \PT-symmetric extensions. 
These include the massive Thirring and Gross-Neveu models as well as 
 a novel spinor model 
that we introduce for  comparison purposes.  Like its two rivals, the new model is U(1)-invariant, Lagrangian and 
has a cubic nonlinearity.

We will show that  
 the  \PT-symmetric extensions of the three models
have different numbers of conservation laws --- from infinitely many to none. 
Despite this difference in the regularity of the dynamics, 
the three \PT-symmetric models will prove to be
 surprisingly coherent as far as their localised solutions are concerned.

The paper is organised as follows.
Section \ref{linear_PTD} provides a motivation for our choice of the \PT-symmetric extension of the linear Dirac equation. 
Having demonstrated that  the  plain addition of gain to one mode and loss to the other leads to an unstable equation,
we select a 
particular stable variant of the \p and \T-breaking perturbation.   This perturbation can also be interpreted as a
balanced gain and loss --- yet for a different pair of modes.

In section \ref{Nonlinear}, we identify a two-parameter family of cubic spinor systems that are invariant under  the full
Lorentz group and the U(1) phase transformations, consider its \PT-symmetric extension and choose three representatives of this family. These are  the massive  Thirring
and 
 Gross-Neveu models, as well as a new simple spinor equation with cubic nonlinearity. 

The physical sources of the \PT-symmetric perturbations
of these spinor systems  are elucidated in section \ref{origin}. 
Another aim of that section is to illustrate the occurrence of three particular types of 
cubic nonlinearity in simple model settings.

The \PT-symmetric Thirring model is scrutinised in section \ref{Thirring}.
We establish that this nonlinear Dirac equation is gauge-equivalent to the original (``parent")  massive Thirring model.
Accordingly,   the \PT-symmetric model represents a completely integrable system and has infinitely many conserved quantities. 
We derive, explicitly,  the first three of these. An explicit expression for the \PT-symmetric soliton  is also produced
in that section.

The following section focusses on the \PT-symmetric Gross-Neveu equation. 
Here our contributions include (a) the local momentum conservation law and (b) an 
exact explicit soliton solution for this model. 

In sections \ref{new_mod} and \ref{new_mod_gamma} we derive an exact soliton solution for our novel spinor model. 
Section \ref{new_mod} deals with the parent system (the equation with no gain and no loss). In this case the soliton is obtained in explicit form.
The subsequent section considers the \PT-symmetric extension of the model; here the 
solution is obtained as a quadrature. The stability of the novel solitons is examined in section \ref{stability}.

Finally, section  \ref{Conclusions} compares all three spinor models and draws general conclusions.

\section{\PT-symmetric extension of the Dirac equation} 
\label{linear_PTD}

The covariant form of 
the linear Dirac equation in the free space is
\be
i \gamma^\mu \partial_\mu \psi + \psi=0.
\label{A1}
\ee
Here
$\partial_\mu = \partial/ \partial x^\mu$, where  $x^\mu$ is the space-time two-vector,
with $x^0=t$ denoting the temporal and $x^1=x$ spatial coordinate.
In   $\gamma^\mu \partial_\mu$,
the Einstein summation convention is implied: $\gamma^\mu \partial_\mu= \gamma^0 \partial_0  + \gamma^1 \partial_1$.
The $\psi$ is a  Lorentz spinor, 
\[
\psi=   \left(  \begin{array}{c} u \\ v \end{array} \right),
\]
where the components $u(x,t)$ and $v(x,t)$ are complex,
and $\gamma^\mu$ are the Dirac $\gamma$-matrices.
We use the following representation for the $\gamma$-matrices:
\be
\gamma^0= \left(\begin{array}{cc} 0 & 1 \\ 1 & 0 \end{array} \right),
\quad
\gamma^1=  \left(\begin{array}{cc} 0 & 1 \\ -1 & 0 \end{array} \right).
\label{Dg}
\ee

When written in components, the 
 Dirac equation has the form of a system
\be
i (u_t-u_x)  + v =0,
\quad 
i  (v_t+v_x)+  u=0,
\label{A00}
\ee
or simply 
\be
i u_\xi  + v =0,
\quad 
i  v_\eta+  u=0,
\label{A0}
\ee
where we have introduced the light-cone coordinates
\[
\eta= \frac{t+x}{2}, 
\quad
\xi=\frac{t-x}{2}.
\]
In terms of the light-cone variables, the proper Lorentz transformation has the form 
\be
\eta \to      e^\beta        \eta,
 \quad
\xi \to   e^{-\beta}    \xi,
\label{A2} 
\ee
where $\beta$ is a real boost parameter. 
(The velocity of the moving reference frame is $\tanh \beta$.)
The components of the spinor $\psi$ transform as
\be
u \to  e^{-\beta/2}  u, 
\quad 
v \to  e^{\beta/2}  v.
\label{A3}
\ee

As one can readily check, the system (\ref{A0})  is invariant under the
 spatial reflections 
  \be
 \mathcal P: x^1 \to -x^1, 
 \quad u \to v, 
 \quad v \to u,
 \label{Ptr}
 \ee
  time reversals
 \be
 \mathcal T: x^0 \to -x^0, 
 \quad u \to v^*, 
 \quad v \to u^*,
 \label{Ttr}
 \ee
 and the proper Lorentz transformations (\ref{A2})-(\ref{A3}).
 Our aim is to identify a physically meaningful and 
 mathematically consistent 
  \PT-symmetric extension of the Dirac equation.  
  The extended system must be  invariant under the combined $\mathcal{PT}$ transformations --- but not under the $\mathcal P$- or $\mathcal T$-reflections individually.
  The physical requirement is that  it remain
  invariant under the  Lorentz boosts (\ref{A2})-(\ref{A3})
   and the U(1) rotations $u \to u e^{i \theta}$, $v \to v e^{i \theta}$ with real constant $\theta$.
  The added perturbation terms are expected to admit the gain-and-loss interpretation --- in some physical contexts,
   at least.
Mathematically, the equation with small perturbation (small gain-loss coefficient)
  should be stable;  that is,  all its solutions with initial conditions satisfying
  \[
  |u(x,0)|^2+|v(x,0)|^2<C   \quad (-\infty<x<\infty)
  \]
  with some $C>0$, 
 should remain bounded in some norm as  $t \to \infty$.

The pair of equations \eref{A00}  bears some similarity with 
 a vector Schr\"odinger equation 
 \be
 iu_t +u_{xx}  + v =0,    \quad
 iv_t+ v_{xx} +  u=0,
 \label{dimer} \ee
 which is commonly used  as a model 
  of the diffractive waveguide coupler  \cite{Akhmediev}.
 The \PT-symmetric extension of the  system \eref{dimer}
 describes the coupler with gain and loss  \cite{NLS_dimer}:
  \begin{eqnarray}
   iu_t+ u_{xx} +  v  & =   i \gamma u,  \nonumber   \\
 iv_t +v_{xx}  + u   &= - i \gamma v. \label{dgl}
 \end{eqnarray} 
 Here $\gamma$ 
 is a positive gain-loss coefficient, $0 \leq \gamma \leq 1$ (not to be confused with the $\gamma$-matrices in \eref{A1} and \eref{Dg}).
 The system \eref{dgl} is invariant under the \PT \/ transformation where the \p-operator is given by \eref{Ptr} and 
  \T  has the form 
   \[
 \mathcal T: t \to -t, 
 \quad u \to u^*, 
 \quad v \to v^*.
\]
 (This transformation  is different from  \eref{Ttr} but still acceptable 
  because $u$ and $v$ are not required to transform as  components of  a spinor in \eref{dimer}-\eref{dgl}.)

Modelling on the Schr\"odinger dimer \eref{dgl}, one could add
gain and loss to the Dirac system \eref{A00}:
 \begin{eqnarray}  
 i u_t-iu_x + v & = i \gamma u,    \nonumber \\
i v_t+iv_x + u   & = -i \gamma v.   \label{ill} 
\end{eqnarray}
However, unlike the Schr\"odinger system \eref{dgl}, 
 this   ``naive" \PT-symmetric extension turns out to have an unstable dispersion relation:
\[
\omega^2= 1 +(k-i \gamma)^2.
\]
The instability is caused by the disbalance between gain and loss in \eref{ill};
indeed, \eref{ill} is not \PT \/ invariant under the spinor transformations \eref{Ptr}-\eref{Ttr}. 

There is a whole range of \PT-symmetric perturbations of the system \eref{A00}
with stable dispersions --- for example, the system
\begin{eqnarray}
  iu_t- i u_x +  v  &  =   i \gamma u_x,  \nonumber \\
 iv_t +i v_x  + u  &  =  i \gamma v_x
 \label{J1}
\end{eqnarray} 
with the dispersion relation
\be
(\omega+ \gamma k)^2 = 1+k^2,     \label{d1}
\ee
or a pair of equations 
\begin{eqnarray} 
  iu_t- i u_x +  v  & =   \gamma u,  \nonumber \\
 iv_t +i v_x  + u  & =  - \gamma v,
 \label{J2}
\end{eqnarray} 
with
\be
\omega^2= 1+ (k+\gamma)^2.    \label{d2}
\ee
While these \PT-symmetric systems may be of interest in some physical contexts, in
 the present study we are focussing on a different extension of the Dirac equation:
\begin{eqnarray}  
 i (u_t-u_x) + v  &=  \gamma v,     \nonumber \\
i(v_t+v_x) + u   & = - \gamma u.
\label{A5}
\end{eqnarray}
The corresponding dispersion relation is
\be
\omega^2=1-\gamma^2 +k^2.  \label{d3}
\ee
Unlike equations \eref{d1} and \eref{d2}, the relation \eref{d3} exhibits the symmetry breaking 
as $\gamma$ exceeds $\gamma_c=1$.  This suggests that  the $\gamma$-terms in \eref{A5} may
account for the gain and loss of energy.

This conjecture turns out to be indeed correct.
Defining 
\be
u_1= \frac{u-iv}{2},
\quad
u_2=\frac{v-iu}{2},
\label{A62}
\ee
equations 
\eref{A5} are transformed into
\begin{eqnarray}
iu_{1t} +u_{2x} +u_2   & = i \gamma u_1, 
\nonumber \\
iu_{2t}- u_{1x} +u_1       &  =- i \gamma u_2.
\label{A640}
\end{eqnarray}
According to the representation \eref{A640}, the system comprises two interacting modes,
where
the $u_1$ mode is gaining and $u_2$ losing energy 
at an equal rate $\gamma$. 
Therefore equation \eref{A5} is more likely to occur in a situation of the 
balanced pump and dissipation than equations \eref{J1} or \eref{J2}. 

Another reason for favouring the extension \eref{A5} over \eref{J1} and \eref{J2}, 
is that the \PT-symmetric terms in \eref{A5} preserve and those in \eref{J1}, \eref{J2} break
the Lorentz invariance of the Dirac equation \eref{A00}. This can be readily verified by changing  to the light-cone variables and 
using the transformation rules \eref{A2}-\eref{A3}.

The model \eref{A5} has originally appeared in \cite{Bender_spinor} 
as a \PT-symmetric free-fermion quantum theory, outside the gain and loss context. 
We should also note two recent publications \cite{Cuevas} and \cite{Sakaguchi}
where a closely related \PT-symmetric extension of the Dirac equation was introduced.

 For the sake of completeness, we need to mention another stable and \PT-symmetric extension 
 of \eref{A00}:
 \[
 i (u_t-u_x) + v  =   \gamma v,     \quad
i(v_t+v_x) + u    =   \gamma u,
\]
with the dispersion $\omega^2= (1 - \gamma)^2+k^2$.
This system is  reducible to \eref{A00} by a trivial coordinate scaling 
and its $\gamma$-terms do not have any gain and loss interpretation.

\section{Nonlinear spinor models} 
\label{Nonlinear}   

Turning to the nonlinear Dirac equations, we restrict ourselves to considering the simplest, cubic, nonlinearity. 
 The most general cubic Dirac equation that is invariant under the proper Lorentz transformations
and U(1) rotations, has the form
\begin{eqnarray}
iu_\xi+ v + (A uv^*+B vu^*)v=0,   \nonumber \\
iv_\eta+u + ({\widetilde A}  vu^*+ {\widetilde B}  uv^*)u=0,  \label{E1}
\end{eqnarray} 
where $A,B, {\widetilde A}, {\widetilde B}$ are complex parameters. 
If we insist that the system \eref{E1} be invariant under the spatial reflections, 
we will have to let
${\widetilde A}=A$ and $\widetilde B= B$. If the system is required to be invariant under the time reversals,
the parameters should satisfy $\widetilde A=A^*$, $\widetilde B=B^*$. 
Therefore the most general cubic U(1)-symmetric spinor system that is invariant under the full Lorentz group, including 
the \p- and \T-transformations,  has the form \eref{E1} 
with real $A=\widetilde A$ and $B=\widetilde B$.

Adding the $\mathcal{P}$- and  $\mathcal{T}$-breaking terms as in \eref{A5} 
we arrive at the \PT-symmetric perturbation of the general model \eref{E1}:
\begin{eqnarray}
iu_\xi+ (1-\gamma)  v + (uv^*+B vu^*)v=0,   \nonumber \\
iv_\eta+ (1+\gamma) u + (vu^*+ B  uv^*)u=0.  \label{E3}
\end{eqnarray} 
In \eref{E3} we have scaled the parameter $A$ out. This can be done without loss of generality ---
except when $A=0$; the latter case has to be considered separately. 

The above  family of models 
and their soliton solutions is the topic of our interest in this paper. 
We will consider three representatives of this family, 
compare their properties,
derive exact expressions for the solitons and examine the soliton stability.

The first representative of the family \eref{E3} is the \PT-symmetric extension of 
the massive Thirring model, selected by letting $B=0$ in
\eref{E3}:
\begin{eqnarray}
i u_\xi + (1-\gamma) v + u|v|^2=0, \nonumber   \\
iv_\eta+ (1+\gamma) u+ v |u|^2=0.  \label{MTM}
\end{eqnarray}
In the covariant notation, equations \eref{MTM} have the form 
\be
i \gamma^\mu \partial_\mu \psi + (1- \gamma  \gamma^5) \psi  + \frac12  \gamma^\mu       J_\mu       \psi=0.
\label{A6}
\ee
Here $J_\mu$ is a two-vector of current:
$J_\mu= {\bar \psi} \gamma_\mu \psi$; the
 ${\bar \psi}$ stands for the Dirac-conjugate spinor: ${\bar \psi}= \psi^\dagger \gamma^0$,
 and $\gamma^5$ is the matrix defined by
 \[
 \gamma^5=\gamma^0\gamma^1=  \left(\begin{array}{cc} -1 & 0 \\ 0 & 1 \end{array} \right). 
\]
(We alert the reader to a slight abuse of notation here.
In the equation \eref{A6} --- and later in \eref{A7} and \eref{A777} --- we use the 
traditional letters
 $\gamma^\mu$ and $\gamma^5$ for the Dirac  $\gamma$-matrices, whereas $\gamma$ without superscripts is just
a  scalar  --- again, a traditional notation for the gain-loss coefficient in literature on \PT  \/ symmetry.)

Our second nonlinear spinor system 
is the  \PT-symmetric extension of the single-component massive Gross-Neveu model
($B=1$ in the list \eref{E3}):
\begin{eqnarray}
i u_\xi + (1-\gamma) v+ (uv^*+u^*v) v=0,
 \nonumber   \\
iv_\eta+ (1+\gamma)  u+ (uv^*+u^*v) u=0.    \label{GrN}
\end{eqnarray}
The covariant formulation of the Gross-Neveu equation is 
\be
i \gamma^\mu \partial_\mu \psi + (1-\gamma \gamma^5) \psi  +  ({\bar \psi} \psi)       \psi=0.
\label{A7}
\ee

The third \PT-symmetric spinor model we scrutinise in this paper, has the form
\begin{eqnarray}
i u_\xi  + (1-\gamma) v+  u^* v^2=0,  \nonumber   \\
i v_\eta  + (1+\gamma)  u +  v^* u^2=0.
\label{A70}
\end{eqnarray}
This pair of equations results by setting $A=0$ (and scaling $B$ out) in the general system \eref{E1} with real $A=\widetilde A$ and $B=\widetilde B$.
Alternatively, one can write $u=B^{-1/2} \widetilde u$, $v= B^{-1/2} \widetilde v$ and send $B \to \infty$ in the system \eref{E3}.
In this sense, equations \eref{A70} represent the $B=\infty$  special case of \eref{E3}.
When written in covariant notation, the system \eref{A70} reads
\be
i \gamma^\mu \partial_\mu \psi + (1-\gamma \gamma^5) \psi  +  ({\bar \psi} \psi) \psi
- \frac{J_\mu}{2} \gamma^\mu  \psi
=0.
\label{A777}
\ee

As we have 
already mentioned, both the massive Thirring and Gross-Neveu models are utilised, extensively,
in quantum field theory, condensed matter physics and nonlinear optics. 
The novel spinor system, equation \eref{A70},  has the nonlinearity as simple as  Thirring's,
and this fact suggests that it may also find physical applications. 
Below, we derive this system in a simple model context. 

\section{Spinors as amplitudes  for counter-propagating waves}
\label{origin}

In this section we
derive our three nonlinear spinor models as equations for  the amplitudes of the back- and forward-propagating 
waves in a medium that supports waves travelling in both directions. 
The  coupling of the two linear waves is achieved by inserting a grating
(or two gratings)  in the system.
(This is not a unique way to produce the coupling; one could alternatively consider a
time-periodic parameter variation.)
Nonlinear effects also contribute to the coupling.

\subsection{Oscillator lattice with  periodic grating}

As the first prototypical system, we adopt a chain of  oscillators
coupled, symmetrically, to their left and right nearest neighbours.
In the absence of perturbations, this system
only involves second-order time
derivatives and allows waves propagating in either direction. 
We are assuming that the atoms in the chain are moving in the external periodic potential
(a grating)  with the 
period much larger than  the lattice spacing.
In the continuum limit, the above discrete system reduces to the Klein-Gordon equation:
\be
\phi_{tt}+ 2 \sqrt{2}    \gamma \epsilon^2 \sin(2x) \phi_t 
 -\phi_{xx} + \phi
-4 \epsilon^2 
\cos(2x) \phi                 
+  4 \cos(4x) \phi^3=0.
\label{A14}
\ee
The second last term in  \eref{A14} represents the grating, with the wavelength
$\pi$. The first-derivative ($\phi_t$) term looks similar, but it has a different physical meaning. 
This term describes damping with a variable coefficient, changing from positive to negative, and back. 
Both periodic terms are considered to be small perturbations;
accordingly, we have entered a small parameter $\epsilon^2$ in front of each of these.
Finally, the  spatial modulation of the cubic term in \eref{A14}
ensures that 
 the nonlinear terms in the resulting amplitude
equations transform as  the Lorentz spinors.
If we do not enter the $\cos (4x)$ factor, we will end up with a system of amplitude
equations where only the linear part is Lorentz-covariant.
(It is fitting to note that this is not the only way to achieve the covariance. We could have employed
a time-periodic variation of the cubic self-coupling instead.)

We expect the evolution to occur over a hierarchy of 
 space and time scales. Defining
\[
T_n= \epsilon^{2n} t, 
\quad
X_n=\epsilon^{2n} x,
\quad n=0,1,2,..., 
\]
we denote
\[
D_n=\frac{\partial}{\partial T_n}, 
\quad
\partial_n= \frac{\partial}{\partial X_n}.
\]
By the chain rule, 
\be
\partial_t^2= D_0^2+ 2 \epsilon^2 D_0D_1+..., 
\quad
\partial_x^2= \partial_0^2+ 2 \epsilon^2 \partial_0 \partial_1+...
\label{A40}
\ee
Expanding $\phi$ in
powers of the small parameter: \[
\phi=\epsilon \phi_1+ \epsilon^3 \phi_3 +...
\]
and substituting, along with the expansions \eref{A40},  in equation  \eref{A14}, we equate 
coefficients of like powers of $\epsilon$. 

At the lowest, $\epsilon^1$-, order, we have
\[
(D_0^2-\partial_0^2+1) \phi_1=0.
\]
We take the solution in the form of a superposition of two counter-propagating waves with equal wavenumbers:
\be
\phi_1= \frac{1}{\sqrt{3}} \left( u e^{i(\sqrt{2} T_0 +X_0)} 
-v e^{i( \sqrt{2} T_0-X_0)} \right)  + \mathrm{c.c.},
 \label{A51}
 \ee
 where $\mathrm{c.c.}$ stands for the complex conjugate of the preceding terms.
 In \eref{A51},  the amplitudes $u$ and $v$ depend on $T_1, T_2, ...,$ and $X_1, X_2, ...$ 
 --- but not on $T_0$ or $X_0$.
 The factor of $1/\sqrt{3}$ is introduced for later convenience.
 
The order $\epsilon^3$ yields
\be
(D_0^2-\partial_0^2+1) \phi_3 =2F_3,
\label{A52}
\ee
where 
\begin{eqnarray*} 
F_3=  &  (\partial_0\partial_1
-D_0D_1)\phi_1  +2 \cos(2X_0)  \phi_1 
- \sqrt{2}  \gamma   \sin(2X_0) D_0 \phi_1  \\  &  -  2  \cos(4X_0) \phi_1^3.
\end{eqnarray*}
Substituting for $\phi_1$ from \eref{A51}, the right-hand side in \eref{A52} becomes a linear combination
 of resonant and nonresonant harmonics. 
Setting to zero the coefficients of the resonant harmonics gives the system \eref{A70},
\begin{eqnarray}
i u_\xi  + (1-\gamma) v+  u^* v^2=0,  \nonumber   \\
i v_\eta  + (1+\gamma) u +  v^* u^2=0,
\label{A59}
\end{eqnarray}
where
\be
\eta= \frac{T_1/ \sqrt{2} +X_1}{2},
\quad
\xi=\frac{T_1/\sqrt{2}-X_1}{2}.
\label{A54}
\ee

The above analysis  explains how a particular type of cubic nonlinearity in the model \eref{A70} can come into being.
It also  
 illustrates one possible source of the 
\PT-symmetric perturbation term:  a balanced gain and loss
of energy in the system. 

A shortcoming of the above model system \eref{A14} is an artificial implementation of the gain and loss.
It is not immediately clear how the periodically varied damping coefficient in \eref{A14} can be realised physically.

\subsection{Diatomic chain with  periodic coupling}
\label{Diatomic}

To describe a more realistic source of  the \PT-symmetric perturbation,
we turn to a slightly more complex  nonlinear bi-directional medium. This time, 
the system consists of  two chains of oscillators with linear and nonlinear coupling.
A common example of such a system is given by a diatomic chain \cite{diatomic}.
Confining the consideration to the continuum limit, we write
\begin{eqnarray} 
\nonumber 
\phi_{tt} - \phi_{xx}+\phi
-4(1- \gamma) \epsilon^2 \cos(2x) \chi 
+ (A+ 4 B \cos 4x )  \phi \chi^2=0,    \nonumber   \\
\chi_{tt} - \chi_{xx}+\chi
-4(1+ \gamma) \epsilon^2 \cos(2x) \phi     + (A+ 4 B  \cos 4x) \chi \phi^2=0. 
\label{A50}
\end{eqnarray}
Here, the $\epsilon^2 \cos(2x)$ term describes a weak periodic modulation of the 
linear inter-chain coupling. (A temporal variation of the coupling produces an
equivalent set of amplitude equations.)
The two nonlinear terms (proportional to $A$ and $B$, respectively) are introduced for generality.
The coefficients $A$ and $B$ are real; depending on the physical setting, one may choose a particular value for each of these.

Note that the $\chi$- and $\phi$-equations in \eref{A50} have different linear coupling amplitudes, $(1-\gamma)$ vs $(1+\gamma)$. This dissonance
breaks the reflection symmetry between the two chains.
We will show, however, that the asymmetric coupling preserves the \PT-symmetry of the
underlying amplitude equations.

Expanding 
\[\phi= \epsilon \phi_1+ \epsilon^3 \phi_3+ ..., 
\quad
\chi=\epsilon \chi_1+ \epsilon^3 \chi_3 +...,
\]
and substituting in \eref{A14}, we obtain, 
at the lowest order of $\epsilon$:
\begin{eqnarray}(D_0^2-\partial_0^2+1) \phi_1=0, \label{A21} \\
(D_0^2-\partial_0^2+1) \chi_1=0. 
\label{A20} 
\end{eqnarray}
In \eref{A21}-\eref{A20}, we use the notation of the previous  subsection.
We take the solution of 
\eref{A21} describing the wave of the unit wavenumber, travelling to the left:
\be
\phi_1=  u e^{i(\sqrt{2} T_0 +X_0)} + \mathrm{c.c.}
\label{A23}
\ee
The solution of \eref{A20} is taken in the form of a wave travelling to the right:
\be
\chi_1= -v e^{i( \sqrt{2} T_0-X_0)}+  \mathrm{c.c.}
\label{A22}
\ee

The order $\epsilon^3$ yields
\begin{eqnarray}
(D_0^2-\partial_0^2+1) \phi_3 =2F_3,   \label{A26} \\
(D_0^2-\partial_0^2+1) \chi_3 =2G_3,  \label{A25}   
\end{eqnarray} 
where
\begin{eqnarray*}
F_3=   (   \partial_0\partial_1   - D_0 D_1) \phi_1  +2(1- \gamma) \cos(2X_0)  \chi_1   
- (A/2+  2B \cos 4X_0) \phi_1 \chi_1^2,  \\
G_3=   (           \partial_0\partial_1      -  D_0 D_1) \chi_1  +2 (1+ \gamma) \cos(2X_0)  \phi_1 
-  (A/2+ 2 B \cos 4X_0)      \chi_1 \phi_1^2.
\end{eqnarray*}
Substituting for $\phi_1$ and $\chi_1$ from \eref{A23}-\eref{A22},
and setting to zero  coefficients of the resonant harmonics $e^{i(\sqrt{2}T_0+X_0)}$ and
$e^{i(\sqrt{2}T_0-X_0)}$,
we obtain 
\begin{eqnarray}
 i u_\xi + (1- \gamma) v + (A uv^* +B u^*v)  v =0,   \nonumber  \\
  i v_\eta +(1+  \gamma) u + ( A vu^* +B v^*u)u =0,
  \label{A63}
 \end{eqnarray}
 with the light-cone variables as defined in \eref{A54}.

Scaling out the parameter $A$ gives 
 the \PT-symmetric spinor model \eref{E3}.
Therefore the diatomic lattice \eref{A50} with $B=0$ gives rise to the \PT-symmetric Thirring model,
the chain with $A=B$ brings about  the \PT-symmetric Gross-Neveu,
while the equations \eref{A50} with $A=0$ produce the new spinor model \eref{A70}.

Note that the \PT-symmetric perturbation term in \eref{A63} is no longer owing to the damping sign variation.
This time, the $\gamma$-term is due to the  difference in the linear 
coupling amplitudes of the $\phi$ and $\chi$ fields. That is, the breaking of the \p and \T invariances 
 is caused by the coupling asymmetry.

\section{ \PT-symmetric Thirring model}
\label{Thirring} 

Having selected three representative \PT-symmetric spinor models
and exemplified possible sources of the associated nonlinearities, we scrutinise each of the three models individually.
We start with the \PT-symmetric Thirring model, equation \eref{MTM}. 

Defining a new pair of the light-cone coordinates by
\be
\widetilde{\xi}= (1- \gamma) \xi, \quad \widetilde{\eta}=(1+\gamma) \eta,
\label{t1}
\ee
and scaling the components of the spinor as in
\be
\widetilde{u}= \frac{1}{\sqrt{1-\gamma^2}} u,
\quad
\widetilde{v}= \frac{1}{\sqrt{1-\gamma^2}} v,
\label{t2} 
\ee
casts equations \eref{MTM} in the form
\begin{eqnarray}  
 i u_\xi + v  +  (1+\gamma) u|v|^2   &= 0,     \nonumber \\
i  v_\eta + u  + (1-\gamma) v |u|^2 & =0
\label{A10}
\end{eqnarray}
(where we have dropped the tildes). 

The system \eref{A10} has been considered previously \cite{David,BG}.
In particular, the one-soliton solution of \eref{A10} has been obtained \cite{BG}. 
Using the scaling transformation \eref{t1}-\eref{t2} we can readily determine the 
soliton solution of the \PT-symmetric Thirring model \eref{MTM}:
\begin{eqnarray}
u=  \left( \frac{1-\gamma}{1+\gamma} \right)^{1/4}
\frac{\kappa  \, e^{i \gamma \mu(x) }
}{\cosh ( \kappa x + i \alpha )}e^{-i\omega t},
\nonumber  \\
 v=- \left( \frac{1+  \gamma}{1-\gamma} \right)^{1/4} 
 \frac{\kappa  \,  e^{i\gamma \mu(x)}
 }{\cosh ( \kappa x - i \alpha ) } e^{-i\omega t },
 \label{eT}
  \end{eqnarray}
  where 
  \[
  e^{i \mu(x)}= \frac{ \cosh ( \kappa x - i \alpha )} 
  { \cosh (  \kappa x + i \alpha )},
  \quad
  \omega= \sqrt{1-\gamma^2} \cos (2 \alpha),
  \quad
  \kappa= \sqrt{1-\omega^2-\gamma^2},
    \]
  and $\alpha$ is a free parameter, $0 < \alpha < \frac{\pi}{2}$. 
  
  It is fitting to note that the general $N$-soliton solution of the system \eref{A10} is also available in literature. 
This solution is expressible in terms of  determinants of $N \times N$ and $(N+1) \times (N+1)$  matrices \cite{BG}.
  Using the scaling  \eref{t1}-\eref{t2} it is straightforward to obtain the corresponding  explicit $N$-soliton 
 solution of the \PT-symmetric model \eref{MTM}.

The \PT-symmetric extension \eref{MTM} is gauge-equivalent to the original (``parent")
 Thirring model.
Indeed, the following local conservation law follows from  \eref{A10}:
\be
 \partial_\xi |u|^2 +     \partial_\eta |v|^2 =0.
\label{A11}
\ee
Equation \eref{A11} implies that there exists a potential $W(\eta, \xi)$ such that 
\[
|u|^2=  -\partial_\eta W,
\quad
|v|^2=   \partial_\xi W.
\]
The gauge transformation \cite{BG}
\be
u= e^{i \gamma W} \hat{u}, 
\quad
v= e^{i \gamma W} \hat{v}
\label{t3}
\ee
takes \eref{A10} to the ``parent" Thirring model (equation \eref{MTM} with $\gamma=0$):
\begin{eqnarray}
 i \hat{u}_\xi + \hat{v}  +   \hat{u}|\hat{v}|^2   &= 0,     \nonumber \\
i  \hat{v}_\eta + \hat{u}   +  \hat{v} |\hat{u}|^2 & =0.
\label{A12}
\end{eqnarray}

Therefore, the \PT-symmetric extension of the Thirring model, 
equation \eref{MTM}, is a completely integrable system --- like the original Thirring model itself \cite{Mikhailov}.
This implies, in particular, that equation \eref{MTM} has infinitely many functionally-independent 
conserved quantities. 

The physically-meaningful conservation laws are the electric charge conservation
\be
q_t+ j_x=0,
\label{ch}
\ee
 where
\begin{eqnarray}
q=|u|^2+|v|^2- \gamma (|v|^2-|u|^2), \quad
  j= |v|^2-|u|^2-\gamma(|u|^2+|v|^2);
\end{eqnarray}
energy conservation
\be 
\mathcal H_t +  \mathcal J_x=0,
\label{ener}
\ee
 where
\begin{eqnarray}
\mathcal H=   &  \frac{i}{2} (u_xu^*-u_x^*u  -v_x v^*+v_x^*v) -  (uv^*+u^*v) -|uv|^2
\nonumber \\
      &   -\frac{\gamma}{2} \left( \frac{|u|^4}{1-\gamma} -\frac{|v|^4}{1+\gamma}\right),
\nonumber \\
\mathcal J  =  &   \frac{i}{2} (u_t^*u-u_t u^* +v_tv^*-v_t^*v) +  \gamma (uv^*+u^*v)     \nonumber \\ 
&      + \frac{\gamma}{2} \left( \frac{|u|^4}{1-\gamma} +\frac{|v|^4}{1+\gamma}\right);
\end{eqnarray}
and conservation of momentum 
\be 
\mathcal P_t + \Phi_x=0,
\label{mom}
\ee
 where
\begin{eqnarray}
\mathcal P   = &  \frac{i}{2} (u_xu^*-u_x^* u +v_xv^* -v_x^* v) -\frac{\gamma}{2} \left( \frac{|u|^4}{1-\gamma} + \frac{|v|^4}{1+\gamma} \right),              
\nonumber \\
\Phi =  &   \frac{i}{2} (u_t^*u -u_tu^* + v_t^*v - v_tv^*)   -(1- \gamma) (uv^*+u^*v)      -|uv|^2        \nonumber \\  &    +\frac{\gamma}{2} \left( \frac{|u|^4}{1-\gamma} - \frac{|v|^4}{1+\gamma} \right).                                  
\end{eqnarray}

Note that the above conservation laws cannot be established using the Noether theorem as
the \PT-symmetric Thirring model \eref{MTM} does not admit a Lagrangian in its $u$ and $v$ variables.
We have obtained \eref{ch}, \eref{ener} and \eref{mom} 
 by means of the 
gauge transformation \eref{t3} and scaling \eref{t1}-\eref{t2} from the corresponding conservation laws of the original Thirring model.

We close this section with a comment on stability. 
The soliton of the original Thirring model \eref{A12} is  stable  due to the complete integrability of that equation.
(For rigorous stability analysis, see \cite{Dima}.)
In view of the gauge equivalence of \eref{A10} and \eref{A12},  the soliton \eref{eT} of the extended model is also linearly and nonlinearly stable
--- regardless of the gain-loss coefficient $0 \leq \gamma <1$ and frequency $-\sqrt{1-\gamma^2}  < \omega < \sqrt{1-\gamma^2}$.

\section{\PT-symmetric Gross-Neveu model}
\label{GNmodel}

The \PT-symmetric extension of the Gross-Neveu model, equation \eref{GrN}, 
was introduced in \cite{Cuevas} (though in a different formulation). 
Like the present study, the earlier investigation focussed on solitons.

\subsection{Explicit soliton solution}

The soliton solution of the original ($\gamma=0$) Gross-Neveu model is known explicitly  \cite{LKG}.
Using a Newtonian path-following algorithm, 
the authors of \cite{Cuevas} continued it  to nonzero $\gamma$ 
and established the  domain of existence of the resulting numerical solution. 
In what follows, we obtain an exact analytical expression  for 
that localised solution.
An analytical  solution has numerous advantages over its numerical approximation;
in particular, 
the domain of existence of  the \PT-symmetric Gross-Neveu  soliton
will be demarcated exactly and explicitly.

Our derivation of the explicit  solution exploits two conservation laws of the
system \eref{GrN}
The authors of Ref \cite{Cuevas} observed  that the equation conserves energy.
 It is not difficult to derive the associated flux in the local conservation law \eref{ener}. We have:
 \begin{eqnarray}
 \mathcal H= \frac{i}{2} (u_xu^*-u_x^* u- v_x v^* + v_x^*v) -(uv^*+u^*v) - \frac12 (uv^*+u^*v)^2,
 \nonumber \\
 \mathcal J=\frac{i}{2} (u_t^*u -u_tu^* -v_t^* v+ v_tv^*) +\gamma (uv^*+u^*v).  \label{GNe}
 \end{eqnarray}
 We also establish the conservation of the field momentum. The momentum 
 density and flux in the local conservation law \eref{mom} have the form
\begin{eqnarray}
\mathcal P= \frac{i}{2} (u_xu^*  -u_x^*u + v_x v^* -v_x^*v) + \gamma (uv^*+u^*v),   \nonumber
\\
\Phi= \frac{i}{2} (u_t^* u-u_t u^*+v_t^* v-v_t  v^*) -  (uv^*+u^*v)  - \frac12 (uv^*+u^*v)^2.  \label{GNp}
\end{eqnarray}

To construct the soliton, we decompose
\be
u(x,t)= a(x) e^{i\theta(x)-i \omega t},
\quad
v(x,t)= - b(x) e^{i \varphi(x)-i\omega t}.
\label{D3}
\ee
Substituting in \eref{GNe}-\eref{GNp}
and assuming that $a(x), b(x) \to 0$ as $|x| \to \infty$, we 
establish two useful relations:
\begin{eqnarray}
a^2= -\frac{(1-\gamma +\sigma/2)\sigma}{2 \omega},   \label{D6}
\\
b^2= -\frac{(1+\gamma +\sigma/2)\sigma}{2 \omega},   \label{D8}
\end{eqnarray}
where 
\[
\sigma(x) = - 2ab \cos (\varphi-\theta).
\]
Taking a product of \eref{D6} and \eref{D8} we obtain
\be
a^2b^2= \left( \frac{\sigma}{2\omega} \right)^2 \left[ \left(1+ \frac{\sigma}{2} \right)^2-\gamma^2 \right].
\label{abl}
\ee
This equation implies that $\gamma^2$ has to be smaller than $(1+\sigma/2)^2$ for all $x$, including $x=\pm \infty$, where $\sigma=0$. 
Therefore, the solution that we are going to construct, will be valid for $\gamma^2 \leq 1$.

Substituting \eref{D3} in \eref{GrN} gives an equation for $a$,
\be
a_x=-(1-\gamma+\sigma) b \sin(\varphi-\theta),
\label{D5}
\ee
a similar equation for $b$, and 
two equations for the phases of the fields:
\begin{eqnarray}
\theta_x= \frac{\sigma/2}{1-\gamma+ \sigma/2} \omega,
\quad
\varphi_x= -\frac{\sigma/2}{1+\gamma+\sigma/2} \omega.
\label{D11}
\end{eqnarray}
(In obtaining \eref{D11}, we made use of \eref{D6} and \eref{D8}.)
Comparing \eref{D11} to \eref{D6} and \eref{D8}, we observe that the function
$\theta(x)$ is monotonically decreasing and $\varphi(x)$ monotonically growing.

Differentiating the relation \eref{D6} in $x$ and using \eref{D5}, 
we arrive at
\be
\sigma_x= 4 \omega ab \sin(\varphi-\theta).  \label{D7}
\ee
This equation implies that regions of growth of $\sigma(x)$ 
correspond to $\omega \sin(\varphi-\theta)>0$
whereas regions of decay are those where $\omega \sin(\varphi-\theta)<0$.

The factor $\sin(\varphi-\theta)$ in the right-hand side of \eref{D7} is determined,
up to a sign, by $\cos(\varphi-\theta)$ which, in turn, can be written as
\[
 \cos(\varphi-\theta)  = -\frac{\sigma}{2ab}.
 \]
With the help of \eref{abl}, this relation gives
\be
\cos^2(\varphi-\theta) =\frac{\omega^2}{(1+ \sigma/2)^2-\gamma^2}.
\label{cvt}
\ee
The numerator in \eref{cvt} needs to be smaller than the denominator for all $x$ --- in particular, for $x=\pm \infty$. 
Accordingly, the parameters $\omega$ and $\gamma$ have to be constrained by $\omega^2+\gamma^2 \leq 1$. 

 Equations \eref{abl} and \eref{cvt} can be used to express the 
right-hand side in \eref{D7} in terms of a single variable, $\sigma(x)$.  
This converts \eref{D7} to a pair of simple separable equations:
\numparts
\begin{eqnarray}
\sigma_x= 
2\sigma \mathcal R, 
\label{R1}  \\
\sigma_x= 
-2\sigma \mathcal R,     
 \label{R2} 
\end{eqnarray}
\label{D10}
\endnumparts 
where
\[
\mathcal R= \sqrt{(1+\sigma/2)^2- \gamma^2-\omega^2}.
\]
One of these equations is valid in the region of growth of $\sigma(x)$ 
and the other one is valid in the complementary region of its decay.

The compatible nonsingular solution of \eref{R1} and \eref{R2},
approaching zero as $|x| \to \infty$, is
\be
\sigma(x)= - \frac{2\kappa^2 }{1+ \rho \cosh (2 \kappa x)},
\label{sig}
\ee
where 
\[
\kappa= \sqrt{1-\rho^2}>0, \quad 
 \rho= \sqrt{\omega^2+\gamma^2}>0.
\]
The absolute values of the $u$ and $v$ components are obtained from \eref{D6}-\eref{D8},
\begin{eqnarray}
a= \frac{\kappa}{\sqrt{\omega}}
\frac{\sqrt{ \rho^2-\gamma +(1-\gamma) \rho \cosh( 2 \kappa x)}}
{1+\rho \cosh( 2 \kappa x)},
\nonumber
\\
b= \frac{\kappa}{\sqrt{\omega}}
\frac{\sqrt{\rho^2+\gamma +(1+\gamma) \rho \cosh( 2 \kappa x)}}
{1+\rho \cosh( 2 \kappa x)}.
\label{AB}
\end{eqnarray}
From equations \eref{AB} it is clear that $\omega$ has to be positive.
Therefore,  the admissible range of $\omega$  is 
$0 <  \omega <  \sqrt{1-\gamma^2}$ 
for each $0 \leq \gamma < 1$.

The phase variables are obtained from \eref{D11}, by integration:
\be
\theta(x) = \theta_0(0)-\theta_0(x),
\quad
\varphi(x)= \varphi_0(x)-\varphi_0(0), 
\label{tvp}
\ee
where
\begin{eqnarray*} 
\theta_0(x)=
\arctan \frac{\rho^2-\gamma+(1-\gamma) \rho e^{2 \kappa x}}
{\omega \kappa},\\
\varphi_0(x)= \arctan \frac{\rho^2+\gamma +(1+\gamma) \rho e^{2 \kappa x}}{\omega \kappa }. 
\end{eqnarray*}
The  solution \eref{sig} satisfies  equation \eref{R1} in the region 
$x<0$ and equation \eref{R2} in the region $x>0$. 
Accordingly, when recovering $\theta(x)$ and $\varphi(x)$ from 
\eref{D11}, the integration constants were chosen so that 
 $ \sin(\varphi-\theta)<0$
 in the region $x<0$ and $ \sin(\varphi-\theta)>0$
 in the region $x>0$.

The soliton of the original Gross-Neveu model ($\gamma=0$) is known to be stable for 
all values of  its frequency $\omega$. For the analytical proof using the Evans function, see \cite{Berkolaiko};
the comprehensive numerical study is in \cite{Num_GN,Taras}. 
The soliton of the \PT-symmetric extension \eref{GrN} was also found to be stable --- for all  $\gamma$ and $\omega$
\cite{Cuevas}. (The analysis of  \cite{Cuevas} appealed to 
 the stability eigenvalues of  the  numerically-determined soliton.) 
Accordingly, we conclude that our explicit soliton solution 
\eref{D3}+\eref{AB}+\eref{tvp} is stable regardless of the gain-loss coefficient $0 \leq \gamma < 1$
 and frequency 
$0 <  \omega <  \sqrt{1-\gamma^2}$.

\subsection{Charge nonconservation} 
\label{chrg}

Despite their common possession of explicit soliton solutions, there is an important difference between 
the \PT-symmetric Thirring and Gross-Neveu models. Whereas the Thirring model has an infinity of 
conservation laws, in the case of the Gross-Neveu equation 
we were  unable to determine 
any other
conserved quantities in addition to  energy and momentum.
 We  therefore {\it conjecture\/} that the system \eref{GrN} with $\gamma \neq 0$ has only two  conservation laws,
 equations \eref{GNe} and \eref{GNp}.

\section{Novel spinor model}
\label{new_mod}

The system \eref{A70} is the  third on our list of spinor models.
Since the model is new, we start with its ``original", i.e. $\gamma=0$, version:
\begin{eqnarray}
i u_\xi  +  v+  u^* v^2=0,  \nonumber   \\
i v_\eta  + u +  v^* u^2=0.
\label{A700}
\end{eqnarray}

The model \eref{A700} admits a Lagrangian, with the density
\be
L=\frac{i}{2} ( u_\xi u^*- u_\xi^*u+ v_\eta v^* -v_\eta^* v)+uv^*+u^*v+ 
\frac{(uv^*)^2+ (u^*v)^2}{2},
\label{newL}
\ee
or, in the covariant formulation, 
\[
L= i {\bar \psi} \gamma^\mu \partial_\mu \psi + {\bar \psi} \psi +
\frac12 ( {\bar \psi} \psi)^2 -  \frac14 J_\mu J^\mu.
\]
The charge, energy and momentum conservation laws are straightforward by means of the Noether theorem.
The local charge is governed by equation \eref{ch} with 
\be
q =  |u|^2+|v|^2,   \quad 
j= |v|^2-|u|^2.
\label{qj}
\ee
The energy and momentum conservation laws have the form \eref{ener} and \eref{mom}, 
where
\begin{eqnarray}
\mathcal H= \frac{i}{2} (u_xu^*- u_x^*u -v_x v^* +v_x^*v)-(uv^*+u^*v) -\frac{ (uv^*)^2+ (u^*v)^2}{2},
\nonumber
\\
\mathcal J= \frac{i}{2} (u_t^*u-u_tu^*+ v_tv^*-v_t^*v),   \nonumber
\\
\mathcal P= \frac{i}{2} ( u_xu^*-u_x^*u+ v_xv^*-v_x^*v),  \nonumber
\\
\Phi= \frac{i}{2} ( u_t^*u-u_tu^* +v_t^*v-v_tv^*) -(uv^*+u^*v) -\frac{(uv^*)^2 + (u^*v)^2}{2}.
\label{PhP}
\end{eqnarray}

Proceeding to the soliton solutions of the 
equations \eref{A700}, we consider stationary (nonpropagating) solitons of the form
\be
u(x,t)= f(x) e^{-i \omega t},
\quad v(x,t)= - g(x) e^{-i \omega t}.
\label{n1}
\ee
The local charge conservation \eref{ch} + \eref{qj}, together with the vanishing boundary conditions
$u,v \to 0$ as $|x| \to \infty$, requires $|f|=|g|$. Hence we write
\be
f(x)=a e^{i \theta},
\quad
g(x)= ae^{i \varphi}.
\label{n2}
\ee
Substituting in \eref{A700} we observe that 
\[
\frac{d}{dx} (\theta+\varphi) =0,
\]
so that  $\theta + \varphi$ is a constant. Using the U(1) invariance of the model we can
set this constant to zero. 
The resulting system has the form
\begin{eqnarray}
a_x=a \sin(2 \theta)- a^3 \sin (4 \theta),  \label{A750}    \\
 \theta_x =\cos(2 \theta) - \omega  -a^2 \cos (4 \theta).
  \label{A75} 
\end{eqnarray}

 The momentum conservation
\eref{mom}+\eref{PhP}, together with the vanishing boundary conditions, furnishes
an invariant manifold of the system 
\eref{A750}-\eref{A75}:
\be
a^2= \frac{2(\cos 2 \theta-\omega)}{\cos 4 \theta}.
\label{A74}
\ee
According to \eref{A74}, $\cos 2 \theta( \pm \infty)=\omega$; hence $|\omega| \leq 1$.
Substituting for $a^2$ in \eref{A75} gives
a simple  separable equation
\be
\theta_x= \omega -\cos(2\theta)
 \label{sep}
\ee
with two solutions,
\be
\theta=\frac{\pi}{2}  + \arctan \left[ \frac{1}{\lambda}
\tanh( \kappa x) \right]
\label{kink}
\ee
and
\be
\theta= - \arctan \left[ \lambda
\tanh( \kappa x) \right].
\label{antikink}
\ee
Here 
 \[
\lambda= \sqrt{\frac{1-\omega}{1+\omega}},
\quad 
\kappa= \sqrt{1-\omega^2}.
\]

The above solutions should be filtered using the equation 
\be
a^2= -\frac{2\theta_x}{\cos 4 \theta},
\label{asq}
\ee
 which is a  consequence of  \eref{A74} and \eref{sep}.
Starting with the  monotonically growing solution  \eref{kink}, we observe that 
 the corresponding $\cos 4 \theta$ passes through $1$
as $x$ goes through the origin.
According to \eref{asq}, this solution produces  negative $a^2$ and should be discarded.

On the other hand, the
 function  \eref{antikink}
  is monotonically decreasing from $\arctan \lambda$
to $-\arctan \lambda$. The right-hand side of  \eref{asq}  is nonnegative
provided 
 $ |4 \theta| \leq \pi/2$. Therefore $\arctan \lambda$ must not exceed $\pi/8$.
 This gives the range of admissible
 $\omega$:
\[
\frac{1}{\sqrt{2}} \leq  \omega \leq 1.
\]

The absolute value of the $u$ and $v$ components is determined from \eref{A74} and \eref{antikink}:
\be
a= \sqrt{2(1-\omega)} \, \sech(\kappa x)
\left[ \frac{1+ \lambda^2 \tanh^2 (\kappa x)}
{1- 6 \lambda^2 \tanh^2 (\kappa x)+ \lambda^4 \tanh^4 (\kappa x)}
\right]^{1/2}.
\label{A76}
\ee

The function \eref{A76} with $\omega$ in the range $\frac34  \leq \omega < 1$
is unimodal (bell-shaped). 
In the remaining part of the parameter interval, $\frac{1}{\sqrt 2}< \omega < \frac 34$, the function $a(x)$ has two
humps placed at $x=\pm x_m$, where $x_m$ is the positive root of
\[
\tanh^2 (\kappa x_m)= \frac{1+\omega}{1-\omega} \times
\frac {1-\omega -\sqrt{\omega^2-1/2}}{1+\omega +\sqrt{\omega^2-1/2}}.
\]
As $\omega \to 1/\sqrt{2}$, the two humps diverge to infinities: $x_m \to \infty$.

Equations \eref{n1}, where $f=ae^{i \theta}$ and $g=ae^{-i \theta}$, while  $a$ and $\theta$ are as in \eref{A76}
and
  \eref{antikink}, provide an explicit soliton solution to the new spinor model \eref{A700}.

\section{\PT-symmetric extension of the new  model:  solution by quadrature}
\label{new_mod_gamma}

Finally, we consider
the \PT-symmetric extension of the model \eref{A700}:
\begin{eqnarray}
i(u_t-u_x) + (1-\gamma) v + v^2u^* &=0,  \nonumber  \\
i(v_t+v_x) +(1+\gamma)u+ u^2 v^*  & =0.  \label{F1} 
\end{eqnarray}

Neither the local charge   \eref{ch}+\eref{qj}
nor the energy-momentum conservation laws
\eref{ener}+\eref{PhP} and \eref{mom}+\eref{PhP} persist as $\gamma$ is taken away from zero. 
In fact we were unable to establish {\it any\/} conservation laws for this \PT-symmetric spinor equation. 
We conjecture that there aren't any.

The lack of conservation laws deprives us of prior knowledge of the invariant manifold that harbours the homoclinic trajectory
in the four-dimensional phase space of the  stationary system. 
As a result, we will be able to construct the exact soliton solution as a quadrature --- but not explicitly.

\subsection{Invariant manifold} 

Assuming stationary solutions of the form \eref{n1} and
letting
\be
f=a(x) e^{i  \theta(x)},
\quad
g= b(x) ^{i  \varphi(x)},
\label{X1} 
\ee
equations  \eref{F1} reduce to
 a four-dimensional stationary  system
\begin{eqnarray}
a_x= -(1-\gamma)b \sin \alpha + ab^2 \sin 2 \alpha,  \label{S20}
\\
b_x= -(1+\gamma) a \sin \alpha+ ba^2 \sin 2 \alpha,  \label{S21} 
\\
\theta_x= -\omega +(1-\gamma) ba^{-1} \cos \alpha -b^2 \cos 2 \alpha,  \label{S22} 
\\
\varphi_x= \omega- (1+\gamma) ab^{-1} \cos \alpha + a^2 \cos 2 \alpha, \label{S23}
\end{eqnarray}
where 
\[
\alpha=\varphi-\theta.
\]
Here we assume that  $a(x), b(x) \to 0$ as $|x| \to \infty$.

The system \eref{S20}-\eref{S23} can be conveniently analysed 
using the Stokes vector ${\vec \mathcal R}= (X,Y,Z)$, where
\begin{eqnarray}
X= 2 ab \cos \alpha,   \quad  &
Y= 2ab \sin \alpha,   \\
Z= a^2-b^2, 
\quad    &
\mathcal R=  |{\vec \mathcal R}|= a^2+b^2.
\end{eqnarray}
(For  review and references on the Stokes coordinates see \cite{BPD}.)
When transformed to the Stokes variables,
equations \eref{S20}, \eref{S21} and a 
combination of \eref{S22}-\eref{S23} 
form a self-contained dynamical system 
 in three dimensions:
\begin{eqnarray}
{\dot X }=(  \mathcal R - 2\omega) Y,      \label{S1}
\\
{\dot Y}= 2 \omega X +(X-2) \mathcal R - 2 \gamma Z,  \label{S2}   \\
{\dot Z}= 2 \gamma Y.    \label{S3}   
\end{eqnarray}
Here the overdot indicates derivative w.r.t. $x$.

Another linear combination of \eref{S22} and \eref{S23} 
is a stand-alone equation for the variable 
$ \beta= \theta+\varphi$:
 \be
{\dot  \beta}  = 2X  \frac{ Z(X-1) -\gamma \mathcal R}{\mathcal R^2-Z^2} -Z.
 \label{beta}
 \ee
Once $X,Y$ and $Z$ have been determined, $\beta$ can be obtained from \eref{beta} by simple integration.

The quantity $\mathcal R$ is governed by 
equation 
\be
{\dot {\mathcal R}}= 2 (X-1) Y,    \label{S4}
\ee
which is a consequence of \eref{S1}-\eref{S3}.
From  \eref{S1} and \eref{S4} we obtain a simple separable equation
\be
\frac{dX}{d \mathcal R} = \frac{\mathcal R- 2 \omega}{2(X-1)}.
\label{S7}
\ee
The solution curve satisfying the initial condition $\left. X \right|_{\mathcal R = 0}=  0$,
is
\be
\frac12 (\mathcal R- 2 \omega)^2 -(X-1)^2= 2 \omega^2-1.
\label{S5}
\ee
Equation \eref{S5} gives an invariant manifold of the system \eref{S1}-\eref{S3}
containing the homoclinic trajectory that we are trying to determine.

\subsection{High-frequency soliton ($\omega^2 > 1/2$)} 
\label{HFs}

Assuming, first, that 
\be
2 \omega^2>1,
\label{lb}
\ee
 we define $\rho>0$ such that 
\[
\rho^2= 2 \omega^2-1.
\]
The implicit curve \eref{S5} can be parametrised by letting
\begin{eqnarray}
\mathcal R= 2\omega-\sqrt{2} \rho \cosh \chi,   \label{R} \\
X=1 -  \rho \sinh \chi,
\label{S6}
\end{eqnarray}
where $\chi=\chi(x)$ is a real parameter.
Equations \eref{R} and \eref{S6} constitute one
of the two branches of the hyperbola \eref{S5} --- specifically, the branch with $\mathcal R < 2 \omega$.
(The branch with $\mathcal R > 2 \omega$ is considered  in  subsection  \ref{negfreq}.)

Substituting \eref{R} and \eref{S6} into \eref{S4} we obtain
\be
\label{S10}
{\dot \chi}=      \sqrt{2} Y.
\ee
Comparing this to \eref{S3} gives 
\be
Z=  \sqrt{2} \gamma (\chi-\chi_0),   \label{Z}
\ee
where $\chi_0$ is the value of $\chi$ attained as $|x| \to \infty$:
\be
\cosh \chi_0= \sqrt{2} \, \frac{\omega}{\rho},
\quad
\sinh \chi_0=   \frac{1}{\rho}.
\label{chi00}
\ee
The first equation in \eref{chi00} infers that $\omega$ has to be positive. The second one tells us that $\chi_0>0$.

Differentiating \eref{S10} with respect to $x$ and using \eref{S2} gives an equation of motion of a fictitious classical particle with the coordinate $\chi$:
\be
{\ddot \chi}= - \frac{\partial U}{\partial \chi}.
\label{S11}
\ee
Here the potential $U(\chi)$ can be cast in the form
\be
\label{U}
U=
4\rho^2 \sinh^2 \frac{\chi-\chi_0}{2}  \left( 1-\sinh^2 \frac{\chi+\chi_0}{2} \right) + 2 \gamma^2 (\chi-\chi_0)^2.
\ee
(We have chosen the zero of the potential to be at $\chi_0$: $U(\chi_0)=0$.)
The first integral of  equation \eref{S11} is a sum of  the kinetic and potential energy of 
the  particle:
\be
\frac{{\dot \chi}^2}{2} + U(\chi)=0,
\label{S14}
\ee
where we have taken into account that $\chi=\chi_0$ is an equilibrium.

The potential $U(\chi)$ has a double zero at $\chi=\chi_0$ and two simple zeros, at $\chi_1$ and $\chi_2$.
One can readily check that 
$\chi_0$ is a point of maximum if 
\be
\omega^2+ \gamma^2 <1.   \label{go}
\ee
In what follows we assume that the inequality \eref{go} is satisfied --- for if it were not, 
the particle 
would oscillate about the minimum of $U(\chi)$ and the corresponding solution $\chi(x)$ 
would not  be localised.

Since $U(-\chi_0) >0$, 
 the other zeros are to the left of $\chi_0$ in this case: $\chi_2<-\chi_0< \chi_1< \chi_0$.
 This implies, in particular, that
 \be
 \sinh \chi_1< \sinh \chi_0
 \label{sinh}
 \ee
 and
 \be
 \cosh \chi_1 < \cosh \chi_0.
 \label{ch10}
 \ee
 
  The point $\chi_1$ is 
   to the left of the origin if  $U(0)<0$
  and  to the right of the origin if  $U(0)>0$.
 Here
 \be
 U(0)= 2 \mathrm{sech}^2 \frac{\chi_0}{2} + 2 \gamma^2 \chi_0^2-1.   \label{H11}
 \ee
 It is not difficult to check that when $\gamma$ is smaller than a certain $\gamma_*$, there are $\chi_a$ and $\chi_b$, 
  $0< \chi_a< \chi_b$, such that 
 the quantity \eref{H11} is negative in the interval $\chi_a<\chi_0 < \chi_b$ and positive outside it.
 On the other hand, when $\gamma> \gamma_*$, the expression \eref{H11} is positive for all $\chi_0$. 
 The critical value $\gamma_*$ is given by
 \be
 \gamma_* = \frac12 \mathrm{sech}  \, \mathcal X   \sqrt{ \frac{\tanh \mathcal X}{\mathcal X}}=0.1904,  
 \label{H12}
 \ee
 where $\mathcal X=1.374$ is the  positive root of the equation
 \be
 1+ \mathcal X \tanh \mathcal X= \frac12 \cosh^2 \mathcal X.
 \label{H14}
 \ee
 Since $\omega= \frac{1}{\sqrt 2}  \, \mathrm{coth} \, \chi_0$, we conclude that for small $\gamma < \gamma_*$, there are 
 $\omega_a=\omega_a(\gamma)$ and $\omega_b=\omega_b(\gamma)$, $1/\sqrt{2} < \omega_b <\omega_a$,  such that the point $\chi_1$ lies to the left of the origin 
 if $\omega_b< \omega< \omega_a$ and to the right of the origin if $\omega$ is outside the interval $(\omega_b, \omega_a)$.

There
 is a homoclinic trajectory connecting the saddle $\chi=\chi_0$, ${\dot \chi}=0$ to itself.
The fictitious particle following this trajectory leaves the unstable equilibrium in the infinite past, reaches $\chi=\chi_1$ at ``time" $x=0$,
and returns to $\chi_0$ in the infinite future ($x \to \infty$).

The variable \eref{R} pertaining to this trajectory has one of two possible behaviours depending on the values of $\gamma$ and $\omega$. 
Frequencies  $\omega$ lying outside the interval $(\omega_b,\omega_a)$ correspond to bell-shaped functions $\mathcal R(x)$ decreasing 
from 
\be
\mathcal R(0)= \sqrt2 \rho (\cosh \chi_0-\cosh \chi_1)
\label{F11}
\ee
to zero as $|x|$ changes from 0 to infinity. On the other hand, frequencies satisfying $\omega_b<\omega<\omega_a$  correspond to
bimodal (double-humped) functions $\mathcal R(x)$. As $|x|$ grows from zero in the latter case,  $\mathcal R$ increases from the value \eref{F11} to $\sqrt{2} \rho (\cosh \chi_0-1)$
and only then decays to zero. 
Due to \eref{ch10}, we have $\mathcal R(x)>0$ for all $x$ ($-\infty<x<\infty$); hence $\mathcal R(x)$ 
does represent the magnitude of the vector $(X,Y,Z)$.
Accordingly,  the homoclinic trajectory 
defines  a localised solution of the system \eref{S1}-\eref{S3} for all $\frac{1}{\sqrt 2} < \omega < \sqrt{1-\gamma^2}$.
 The domain of existence on the $(\gamma, \omega)$ plane is illustrated in Fig \ref{existence} (b). 

Note that  $d \chi/d x<0$ in the region $x<0$  and $d\chi/ dx>0$ in $x>0$.
Keeping this correspondence in mind, we  integrate \eref{S14} to obtain
\be
|x |= \int_{\chi_1}^\chi \frac{d\chi}{\sqrt{-2U(\chi)}},
\quad \chi_1 \leq \chi < \chi_0.
  \label{S15} 
  \ee
Equation \eref{S15}  defines the function $\chi(x)$ over the entire real line $-\infty<x<\infty$.
The function $\chi(x)$ is even. As $|x|\to \infty$, we have $\chi(x) \to \chi_0$;
at the origin,   $\chi(0)= \chi_1$. 

Having determined the function $\chi(x)$, equations \eref{R} and \eref{Z} can be used to reconstruct 
the moduli of the $u$ and $v$ components of the spinor soliton:
\begin{eqnarray} 
a(x)  = \sqrt{\omega-\frac{\rho}{\sqrt{2}} \cosh \chi +   \frac{\gamma}{\sqrt{2}} (\chi-\chi_0)},
\nonumber
\\
b(x)= \sqrt{\omega-\frac{\rho}{\sqrt{2}} \cosh \chi - \frac{\gamma}{\sqrt{2}} (\chi-\chi_0)}.
\label{X2}
\end{eqnarray}

To reconstruct the corresponding phase variables, we need to determine their linear combinations, $\alpha$ and $\beta$.
Equation \eref{S6} reads $X=\rho (\sinh \chi_0-\sinh \chi)$;  the inequality \eref{sinh} implies then
 $X(x)>0$ for all $-\infty<x<\infty$.  Because of that,  the angle $\alpha$ can be taken to lie between $-\pi/2$ and $\pi/2$
and we can let $\alpha= \arctan (Y/X)$. Consequently, 
\be
\alpha(x)= \mathrm{sign}  (x) \arctan  \left[  \frac{\sqrt{-U(\chi)}}{\rho  ( \sinh \chi_0 -\sinh \chi)}  \right],
\label{X3}
\ee
where $U(\chi)$ is as in \eref{U}. In the above expression, we have taken into account that 
$Y(x)=\frac{1}{\sqrt 2} \dot \chi$ is positive and negative for $x>0$ and $x<0$, respectively.

The function $\beta(x)$ can be found from \eref{beta} by integration:
\be
\beta(x)= 
- \sqrt2 \gamma \int_0^x   \left[ 
   (\sinh \chi_0- \sinh \chi) Q 
   -   (\chi_0-\chi)
 \right] dx, 
 \label{X4}
 \ee
 where
 \[
Q(\chi) = \rho^2  \frac{
    \cosh \chi_0 -\cosh \chi - (\chi_0-\chi)  \sinh \chi
  }{
\rho^2(\cosh \chi_0- \cosh \chi)^2 - \gamma^2(\chi_0-\chi)^2
}.
\]
Both $\alpha(x)$ and $\beta(x)$ are odd functions, bounded  as $x \to \pm \infty$.

Once $\alpha$ and $\beta$ have been determined, the phases of $u$ and $v$ 
are found simply as
\be
\theta=\frac{\beta-\alpha}{2},
\quad 
\varphi= \frac{\alpha+\beta}{2}.
\label{tbv} 
\ee

\subsection{Solitons with negative frequencies?}
\label{negfreq}

We return to equation  \eref{S5} with $2 \omega^2-1>0$
and consider the second branch  of this hyperbola. Instead of 
 equation \eref{R},  the corresponding $\mathcal{R}$-component is given by
 \be
 \mathcal R= 2 \omega + \sqrt{2} \rho \cosh \chi,
 \label{S30}
 \ee
 while the $X$-component is given  by equation \eref{S6}, as before.
This time, the parameter $\chi$ satisfies 
\be
\label{S101}
{\dot \chi}= - \sqrt{2} Y
\ee
and $Z$ is found to be
\be
Z= -\sqrt{2} \gamma (\chi-\chi_0),   \label{Z1}
\ee
where $\chi_0$ is the value of $\chi$ attained as $|x| \to \infty$:
\be
\cosh \chi_0= -\sqrt{2} \, \frac{\omega}{\rho},
\quad
\sinh \chi_0=  \frac{1}{\rho}.
\label{chi01}
\ee
 The equations  \eref{chi01} imply that $\omega$ has to be negative this time
 while 
 $\chi_0$ remains positive.

 Differentiating \eref{S101} in $x$ and using \eref{S2}, we arrive at the same Newton's equation \eref{S11}
 as in the analysis of the hyperbola branch given by \eref{R}-\eref{S6}. 
 The potential energy of the fictitious particle is given by the same equation \eref{U} as before.
 As we have established, there is a homoclinic trajectory connecting the saddle $\chi=\chi_0, \dot \chi=0$ 
 to itself. (We  assume that inequality \eref{go} is in place.)
 The fictitious particle following this trajectory moves from its equilibrium position at $\chi_0>0$ to the point $\chi_1<\chi_0$ and then returns to $\chi_0$. 
 
This time,  the homoclinic trajectory 
 does not furnish a localised solution of the system \eref{S1}-\eref{S3} though.
 Indeed, in view of the inequality  \eref{ch10}, the variable
 \[
 \mathcal R(x)= \sqrt2 \rho \, \left[ \cosh \chi(x)-\cosh \chi_0 \right]
 \]
 remains negative along the entire trajectory and cannot represent the magnitude of the vector $(X,Y, Z)$.
 We conclude that the system \eref{F1} does not have solitons with frequencies   $\omega < -\frac{1}{\sqrt 2}$.

 \subsection{$\gamma=0$ reduction of quadrature}
 
It is instructive to follow the transformation of the quadrature \eref{S15}-\eref{tbv} 
to the explicit solution \eref{antikink}, \eref{A76}
 once  $\gamma$ is set to zero.
 When $\gamma=0$, we have $Z=0$ so that $a=b$ and
$\cos \alpha= X/ \mathcal R$. With the help of \eref{chi00}, equations
\eref{R} and \eref{S6} give
\be
\cos \alpha=  \frac{1}{\sqrt{2}} \, \mathrm{coth} \, \frac{\chi+\chi_0}{2}.
\label{subs}
\ee
Using \eref{subs} the integration over  $\chi$ in \eref{S15} can be changed to integration over $\alpha$:
\be
2x= \int_0^\alpha \frac{d\alpha}{\cos \alpha -\omega}.
\label{S31}
\ee
In transforming \eref{S15} to \eref{S31}, 
we made use of 
\be
d \chi= \pm 2   
\left[ \sinh^2 \frac{\chi+\chi_0}{2} -1\right]^{1/2}  \sinh \left( \frac{\chi+\chi_0}{2} \right)
  d \alpha.
\label{mp}
\ee
Here the top sign  corresponds to the region $x>0$ (where  $\sin \alpha>0$)
and the bottom sign to $x<0$ (where $\sin \alpha<0$).
We have also used the relation
\[
   \frac{\sinh \frac{\chi_0-\chi}{2}}{\sinh \frac{\chi_0+\chi}{2}}  =   \frac{\sqrt{2}}{\rho}  ( \cos \alpha -\omega).
\]

When $\gamma=0$, we have $\beta=0$ and so $\alpha=-2 \theta$. 
With this observation, 
equation \eref{S31} is nothing but an integral of the separable equation \eref{sep}
with an explicit solution \eref{antikink}.

\subsection{Low-frequency soliton ($\omega^2 < 1/2$)}

We proceed to the situation $2\omega^2<1$ and define $\rho>0$, where
\[
\rho^2=1-2\omega^2.
\]
In this case the hyperbola \eref{S5} 
admits a unique parametrisation 
consistent with the boundary condition $X(\pm \infty)=0$:
\begin{eqnarray}
\mathcal R = 2 \omega   -  \sqrt2 \rho \sinh \chi,   \label{R20} \\
X= 1- \rho \cosh \chi.
\end{eqnarray}
As in   subsection \ref{HFs}, $\chi$ satisfies \eref{S10} and
$Z$ satisfies equation \eref{Z}
where 
 $\chi_0$ is the value of $\chi$ attained as $|x| \to \infty$:
\be
\sinh \chi_0= \sqrt{2} \, \frac{\omega}{\rho},
\quad
\cosh \chi_0= \frac{1}{\rho}.
\label{chi0}
\ee
Equations \eref{chi0} allow both signs of $\omega$: 
positive $\omega$'s correspond to  $\chi_0>0$ and negative
$\omega$'s correspond to  $\chi_0<0$. 

Differentiating \eref{S10} in $x$ we obtain
the Newton's equation \eref{S11} where the potential 
\be
\label{U1}
U=
-4\rho^2 \sinh^2 \frac{\chi-\chi_0}{2}  \left( 2+ \sinh^2 \frac{\chi+\chi_0}{2} \right) + 2 \gamma^2 (\chi-\chi_0)^2.
\ee
The point $\chi_0$ is a point of maximum of $U(\chi)$ if $\omega$ and $\gamma$ satisfy the inequality
\eref{go}. This is the first necessary condition for the existence of the homoclinic orbit.

When $\gamma=0$, 
the point $\chi=\chi_0$ is the only maximum of the potential \eref{U1}.
The potential decreases monotonically in either direction away from $\chi_0$.
Assume  we now keep $\rho$ unchanged and raise $\gamma$.
As  $\gamma$ reaches a certain $\gamma_{\mathrm{max}}>0$,  the potential function develops the second maximum at the point $\chi=\chi_{\mathrm{max}}$.
A simple graphical analysis of the derivative
\begin{eqnarray*}
\frac{dU}{d \chi}= &  -4 \rho^2 \sinh \frac{\chi-\chi_0}{2} 
\left[ \cosh \frac{\chi-\chi_0}{2} + \cosh  \chi \cosh \frac{\chi+\chi_0}{2}  \right]    \nonumber  \\  & + 4 \gamma^2 (\chi-\chi_0)
\end{eqnarray*}
 indicates that  the point $\chi_{\mathrm{max}}$ is on the 
right of $\chi_0$  when $\chi_0<0$ and on the left of $\chi_0$ when  $\chi_0>0$.

As 
$\gamma$ exceeds a critical value $\gamma_c> \gamma_{\mathrm{max}}$, the second maximum reaches above zero: $U(\chi_{\mathrm{max}})>0$. 
In the parameter region $\gamma> \gamma_c$, the function $U(\chi)$ has two simple zeroes in addition 
to the double zero at $\chi_0$: $U(\chi_{1,2})=0$. When $\chi_0<0$, we have 
 $\chi_0< \chi_1 < \chi_2$, and when $\chi_0>0$, the arrangement is  $\chi_2<\chi_1<\chi_0$.
 It is not difficult to realise that in the region $\gamma> \gamma_c$,
  the equation \eref{S11} has a homoclinic orbit.
However, this does not necessarily mean that the three-dimensional dynamical system \eref{S1}-\eref{S3} has one.

Indeed, let $\chi_0<0$. As $x$ changes from minus  infinity to zero,
$\chi(x)$   grows from $\chi_0$ to $\chi_1$. Equation \eref{R20} implies then that  the corresponding $\mathcal R(x)$ 
is negative; this disqualifies the choice $\chi_0<0$.  As a result, there are no solitons with $\omega<0$. 

In contrast, solitons with $\omega>0$ do exist. 
In this case $\chi_0>0$; as $x$ varies from $-\infty$ to $0$,  the parameter $\chi$  decreases from $\chi_0$ to $\chi_1$. 
According to \eref{R20}, the corresponding $\mathcal R(x)$ grows, monotonically,  from $0$ to its maximum value 
\[
\mathcal R(0)= \sqrt{2} \rho (\sinh \chi_0 - \sinh \chi_1)>0.
\]
Therefore   $\mathcal R$ does give the length of the vector $(X,Y,Z)$
and the homoclinic orbit of \eref{S11} defines a localised solution of \eref{S1}-\eref{S3}. 

Note that, unlike the high-frequency solitons considered in section \ref{HFs}, 
the solitons in the range $0< \omega < \frac{1}{\sqrt 2}$ all have a unimodal function $\mathcal R(x)$.

For each $\chi_0>0$, the critical value $\gamma_c=\gamma_c(\omega)$ can be determined
as a root of the system
\[
U(\chi, \gamma)=0, 
\quad
\frac{\partial U(\chi, \gamma)}{\partial \chi}=0.
\]
Substituting from \eref{U1} this system acquires the form 
\begin{eqnarray}
\frac{\gamma^2}{\rho^2} \left( \frac{\xi}{\sinh \xi} \right)^2 = \frac34+ \frac14 \cosh( 2 \xi+2 \chi_0),     \label{W1} \\
\frac{\gamma^2}{\rho^2} \frac{\xi}{\sinh \xi} = \frac34 \cosh \xi+ \frac14 \cosh (3 \xi+ 2 \chi_0),   \label{W2}
\end{eqnarray}
where $\xi=(\chi-\chi_0)/2$. 
Eliminating $\gamma$ 
between \eref{W1} and \eref{W2}, we arrive at a simple
transcendental equation
\be
 F (\xi)= G_\omega(\xi),   
  \label{fg} 
\ee
where
\be
F= \frac{\xi}{\tanh \xi},
\quad
G_\omega= \frac
{3+  \cosh( 2 \xi+2 \chi_0)}
{ 3 +  \cosh (3 \xi+ 2 \chi_0)/ \cosh \xi}.
\label{2f}
\ee 
The subscript $\omega$ in $G_\omega$ serves to remind that 
there is a one-to-one correspondence between $\omega$ and $\chi_0$: $\tanh \chi_0 = \sqrt{2} \omega$.

\begin{figure}[t]
 \begin{center} 
  \includegraphics*[width=0.45\linewidth]{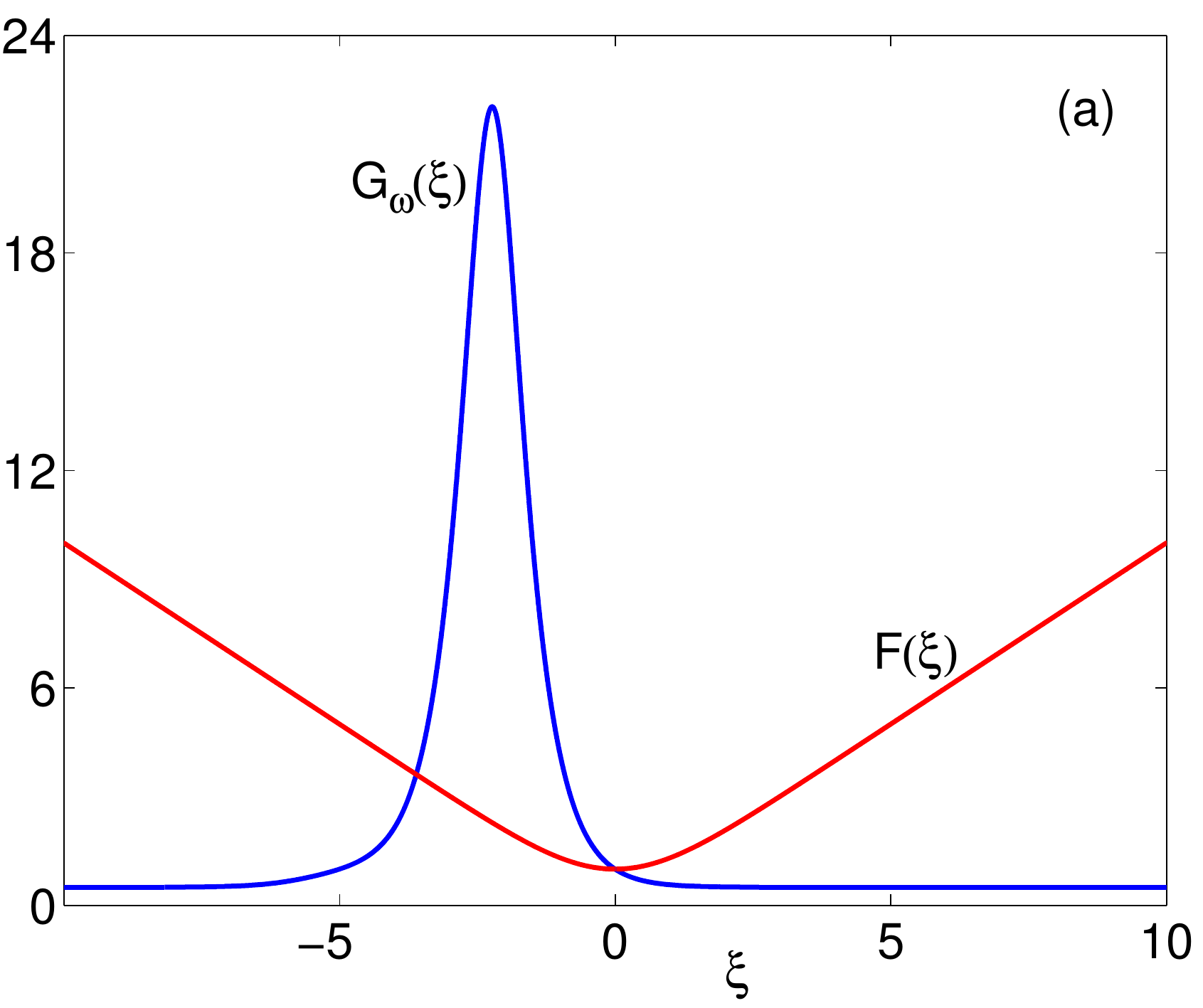} 
  \hspace*{0.05 \linewidth}
   \includegraphics*[width=0.47\linewidth] {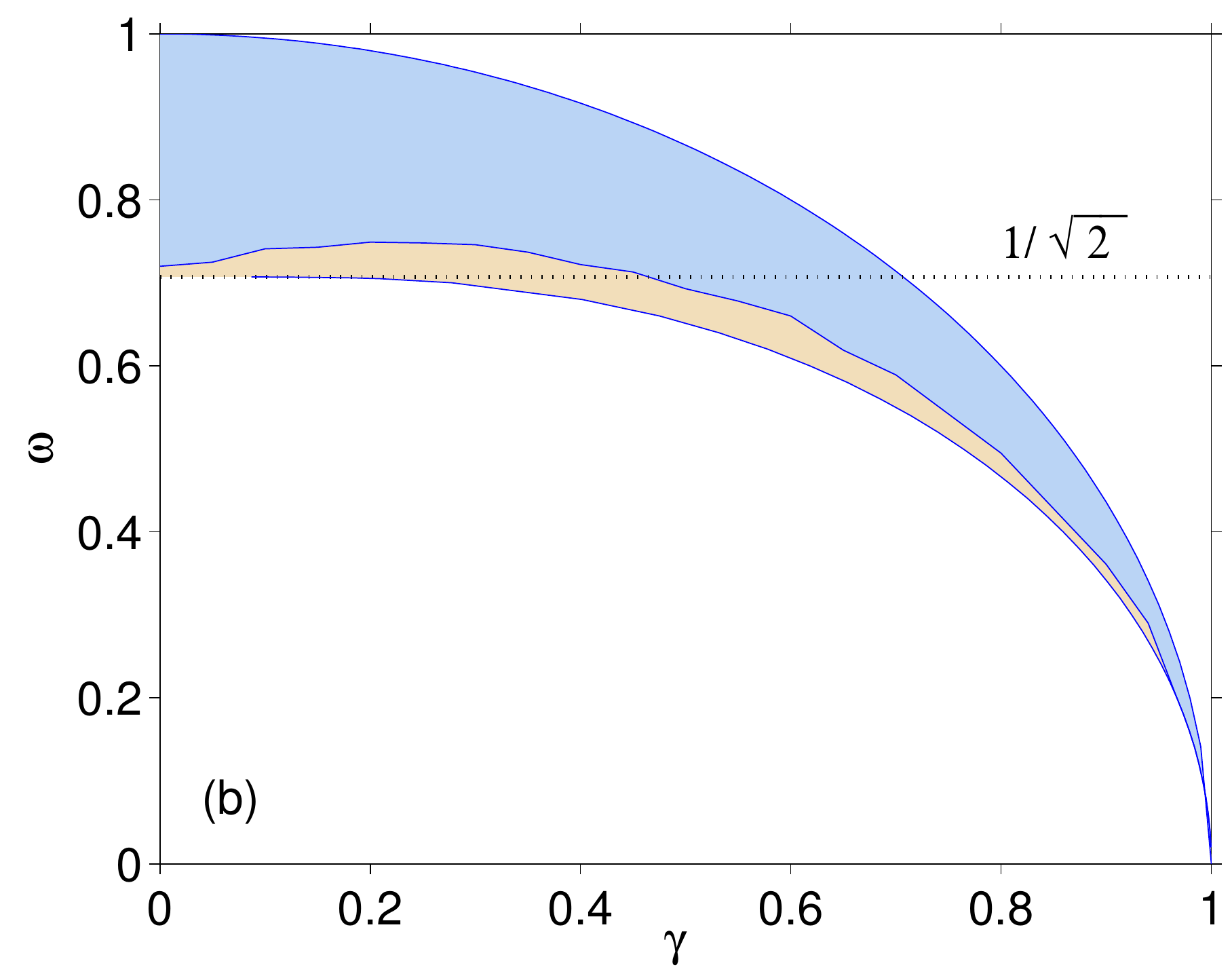} 
  \end{center}
  \caption{\label{existence}
  (a): A graphical solution of the transcendental equation \eref{fg}. 
  Red curve: $F(\xi)$; blue:  $G_\omega(\xi)$.
  In this plot, $\chi_0=5$.
 (b): The domain of existence of the soliton of the model \eref{F1} (coloured blue and brown). 
 The soliton is stable in the subdomain tinted blue while the band of instability is shown in brown.
  The domain  of existence is bounded  by $\omega=\sqrt{1-\gamma^2}$ from above
  and  by $\omega=\omega_c(\gamma)$ from below.
  }
 \end{figure}

The even function $F(\xi)$  has a single extremum (a minimum) 
  at $\xi=0$  and
grows to infinity as
$|\xi| \to \infty$.  The function $G_\omega(\xi)$ also has a single extremum (a maximum)
 at some $\xi<0$,
and approaches $1/2$ as $|\xi| \to \infty$.
The functions $F$ and $G_\omega$ intersect at $\xi=0$ where 
$F(0)=G_\omega(0)=1$. At the point of intersection, $G_\omega(\xi)$ has a negative slope,
\[
\left.  \frac{dG_\omega}{d \xi}  \right|_{\xi=0} = - \frac{\sinh ( 2\chi_0)} {3+ \cosh (2 \chi_0)}<0,
\]
whereas $\left. dF/d \xi \right|_{\xi=0}=0$.
Therefore for each $\chi_0>0$, there is one more intersection, at $\xi=\xi_c <0$.
See Fig \ref{existence} (a). 
The point $\xi_c$ depends on $\omega$.

Having computed $\xi_c$ for a sample of $\omega$ values by means of the
 standard Newtonian iteration,
we use \eref{W1} to determine the corresponding $\gamma_c(\omega)$:
\be
\gamma_c = \rho \frac{\sinh \xi_c}{\xi_c} \sqrt{1+ \frac12 \sinh^2(\xi_c+ \chi_0)}.
\label{W3}
\ee
The inverse function, $\omega=\omega_c(\gamma)$,
 gives the lower boundary of the soliton's domain of existence;
see Fig \ref{existence}(b).

 Note that equation \eref{W3} implies  that $\gamma_c> \rho$.
This furnishes a simple lower bound on the function $\omega_c(\gamma)$:
\[
\omega_c(\gamma)> \frac{1}{\sqrt{2}} \sqrt{1-\gamma^2}.
\]

  Thus we have established  the  domain of existence of solitons with frequencies $\omega<  \frac{1}{\sqrt{2}}$.
  Specifically,   the system  \eref{S1}-\eref{S3}  has a localised solution 
provided $\omega$ lies between $\omega_c(\gamma)$ and $\sqrt{1-\gamma^2}$.
The domain of existence of the low-frequency solitons  seamlessly adjoins the domain
of solitons with $\omega> \frac{1}{\sqrt{2}}$; see Fig \ref{existence} (b).

 The soliton is expressible in terms of the
function $\chi(x)$  given by the quadrature \eref{S15}, with $U(\chi)$  as in \eref{U1}. 
Once the function $\chi(x)$ has been determined, we recover $a(x)$ and $b(x)$ using
\eref{R20} and \eref{Z}:
\begin{eqnarray}
a(x)= \sqrt{\omega-  \frac{\rho}{\sqrt2} \sinh \chi + \frac{\gamma}{\sqrt2} (\chi-\chi_0)},    \nonumber
\\
b(x)= \sqrt{\omega- \frac{\rho}{\sqrt2} \sinh \chi - \frac{\gamma}{\sqrt2} (\chi-\chi_0)}.
\label{X6}
\end{eqnarray}
The function $\alpha(x)$ is determined from 
\be
\alpha(x) = \mathrm{sign} (x) \,  \left[ \arctan  \, \frac {\sqrt{-U(\chi)}}  {\rho(\cosh \chi_0-\cosh \chi)} \right],
\quad
-\frac{\pi}{2}  < \alpha < \frac{\pi}{2},
\label{X7}
\ee
where 
$U$ is as in \eref{U1}.
The function $\beta(x)$ can be found from \eref{beta} by integration:
\be
\beta(x)= 
-\sqrt2 \gamma \int_0^x   \left[ 
   (\cosh \chi_0- \cosh \chi) \widetilde Q 
   -   (\chi_0-\chi)
 \right] dx, 
 \label{X8}
 \ee
 where
 \[
\widetilde Q(\chi) = \rho^2  \frac{
    \sinh \chi_0 -\sinh \chi - (\chi_0-\chi)  \cosh \chi
  }{
\rho^2(\sinh \chi- \sinh \chi_0)^2 - \gamma^2(\chi-\chi_0)^2
}.
\]
Both $\alpha(x)$ and $\beta(x)$ are odd functions, bounded  as $x \to \pm \infty$.
Once $\alpha(x)$ and $\beta(x)$ have been constructed, the phases $\theta(x)$ and $\varphi(x)$ 
are determined from \eref{tbv}.

\section{Stability of solitons in the new model}
\label{stability}

To classify the stability of the soliton of the new spinor model and its \PT-symmetric extension, 
we linearise equations 
 \eref{A700} and \eref{F1} about the 
 solution \eref{n1}. Choosing perturbations of the form 
 \begin{eqnarray*}
 u= \left[ f(x) +z_1(x) e^{\lambda t}  \right]e^{-i \omega t},
 \quad
 v=\left[ -g(x)+z_2(x) e^{\lambda t} \right] e^{-i \omega t}, \nonumber \\
 u^*=\left[ f^*(x) +z_3(x) e^{\lambda t} \right] e^{i \omega t}, 
 \quad
 v^*=\left[ -g^*(x)+ z_4(x) e^{\lambda t} \right] e^{i \omega t}
 \end{eqnarray*}
 gives an eigenvalue problem 
 \be
  H {\bf z} = i \lambda J {\bf z}, \quad
 {\bf z}(\pm \infty)=0.
  \label{EV}
 \ee
 Here
 ${\bf z}=(z_1, z_2, z_3, z_4)^T$;  the operator $H$ is defined by
\begin{eqnarray} 
 H= i  \left( \begin{array}{cc} - \sigma_3 & 0 \\ 0 & \sigma_3 \end{array} \right) \frac{d}{dx}
+
\left( \begin{array}{cc} \omega \sigma_0 + \sigma_1-i\gamma \sigma_2  & 0 \\
0 & \omega \sigma_0 + \sigma_1-i\gamma \sigma_2
 \end{array} \right)    \nonumber \\
 + \left(
 \begin{array}{cccc}
 0 & - 2f^* g & g^2  & 0 \\
 -2 fg^* & 0 & 0 & f^2 \\
 (g^*)^2 & 0 & 0 & - 2 fg^* \\
 0 & (f^*)^2 & -2 f^* g & 0
 \end{array}
 \right),
 \label{St1}
 \end{eqnarray} 
and $J$ is a          diagonal          constant           matrix
\be
J= \left( \begin{array}{cc} - \sigma_0 & 0 \\ 0 & \sigma_0 \end{array} \right).
\label{St2}
\ee
In \eref{St1}-\eref{St2}, $\sigma_{1,2,3}$ are the Pauli matrices, and $\sigma_0$ is the $2 \times 2$ identity matrix. 

The continuous spectrum of $\lambda$ lies on the imaginary axis and consists of two branches.
The first branch has a narrow gap, $| \mathrm{Im} \, \lambda_1| \geq \sqrt{1-\gamma^2} -\omega$.
The second branch's gap is wider: $|\mathrm{Im} \, \lambda_2| \geq \sqrt{1-\gamma^2} + \omega$.

We approximate the boundary conditions in \eref{EV} by ${\bf z}(\pm L)=0$; the bulk of our calculations was done with $L=40$. 
The  Chebyshev differentiation on a nonuniform mesh with $N=1200$ nodes
 converts \eref{EV} to an eigenvalue problem for a $4N \times 4N$ matrix.
The matrix eigenvalues are then computed using a standard numerical routine.

The  soliton solution 
\eref{X2}-\eref{tbv} was examined on a   grid  of $\gamma$ and $\omega$ parameters covering
 the two-dimensional  domain $0< \gamma<1$,
$\frac{1}{\sqrt{2}}< \omega < \sqrt{1-\gamma^2}$. The soliton
\eref{X6}-\eref{X8} was similarly studied by sampling on $0< \gamma<1$ and
$\omega_c(\gamma)< \omega < \frac{1}{\sqrt{2}}$.
We paid special attention to 
the explicit solution 
\eref{antikink}+\eref{A76}, with $\omega$ changing from 1 to $1/\sqrt{2}$. 
The solution with a particular choice of $\gamma$ and $\omega$ was classified as unstable if the spectrum contained eigenvalues with $\mathrm{Re} \, \lambda > 0.003$. 

The soliton's stability properties were found to be qualitatively similar for all $0 \leq \gamma <1$.
Assume the soliton
frequency $\omega$ is decreased while the gain-loss coefficient $\gamma$ is kept constant. 
As $\omega$ passes through a certain $\omega_{\mathrm{inst}}=\omega_{\mathrm{inst}}(\gamma)$,   
a complex quadruplet $\pm \lambda, \pm \lambda^*$ bifurcates from  $\pm i(\sqrt{1-\gamma^2} + \omega)$, the edges
of the wider gap in the continuous
spectrum.
As $\omega$ is further decreased towards $\omega_c(\gamma)$, the lower boundary of the soliton's 
existence domain,
the  real parts of $\lambda$ grow in absolute value (but never exceed $0.3$).
The instability band  $\omega_c(\gamma)< \omega< \omega_{\mathrm{inst}}(\gamma)$   is quite narrow;
see Fig.\ref{existence}(b) where it is tinted brown.

\begin{figure}[t]
 \begin{center} 
   \includegraphics*[width=0.48\linewidth]{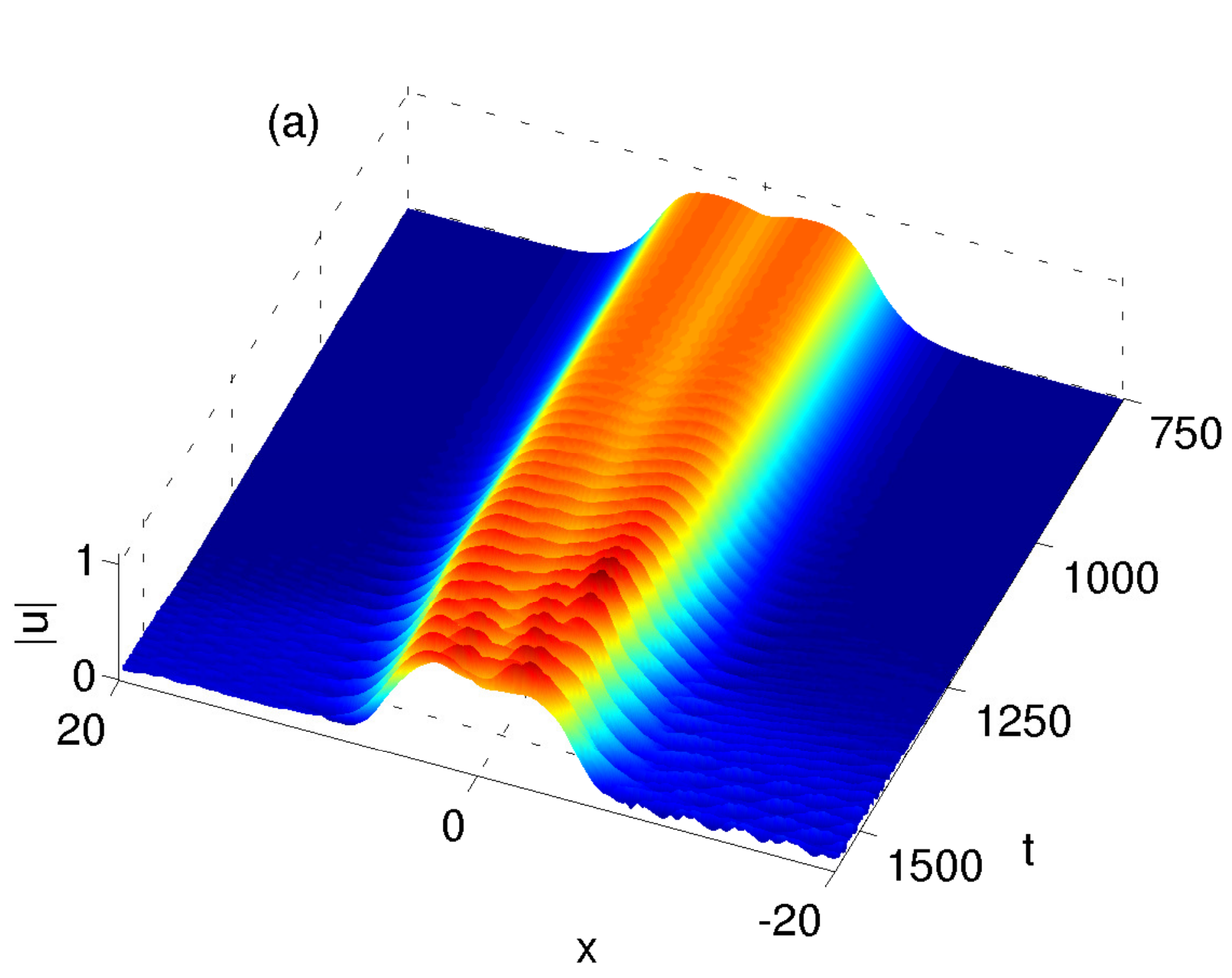} 
  \hspace*{0.02 \linewidth}
   \includegraphics*[width=0.48\linewidth] {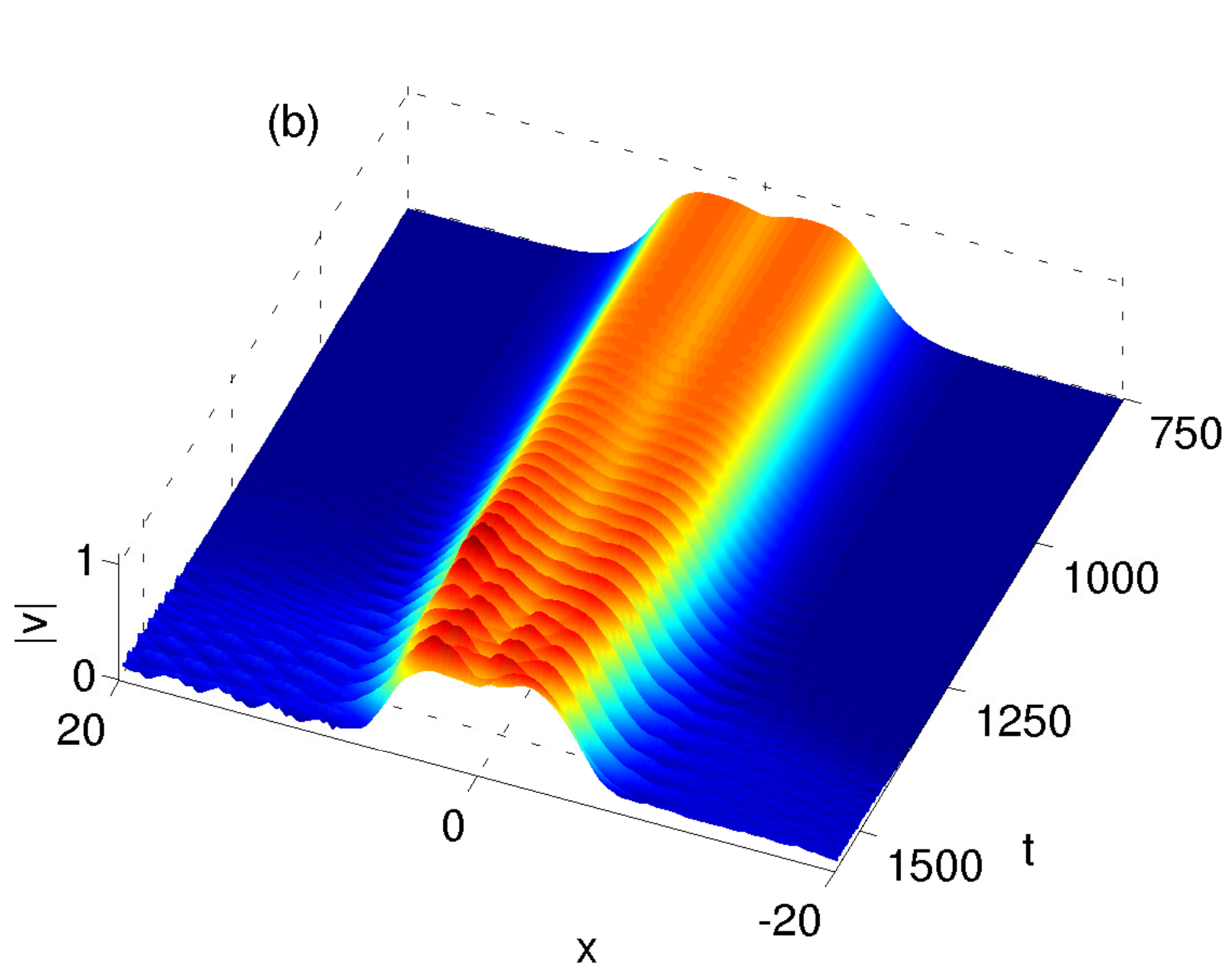} 
  \includegraphics*[width=0.48\linewidth]{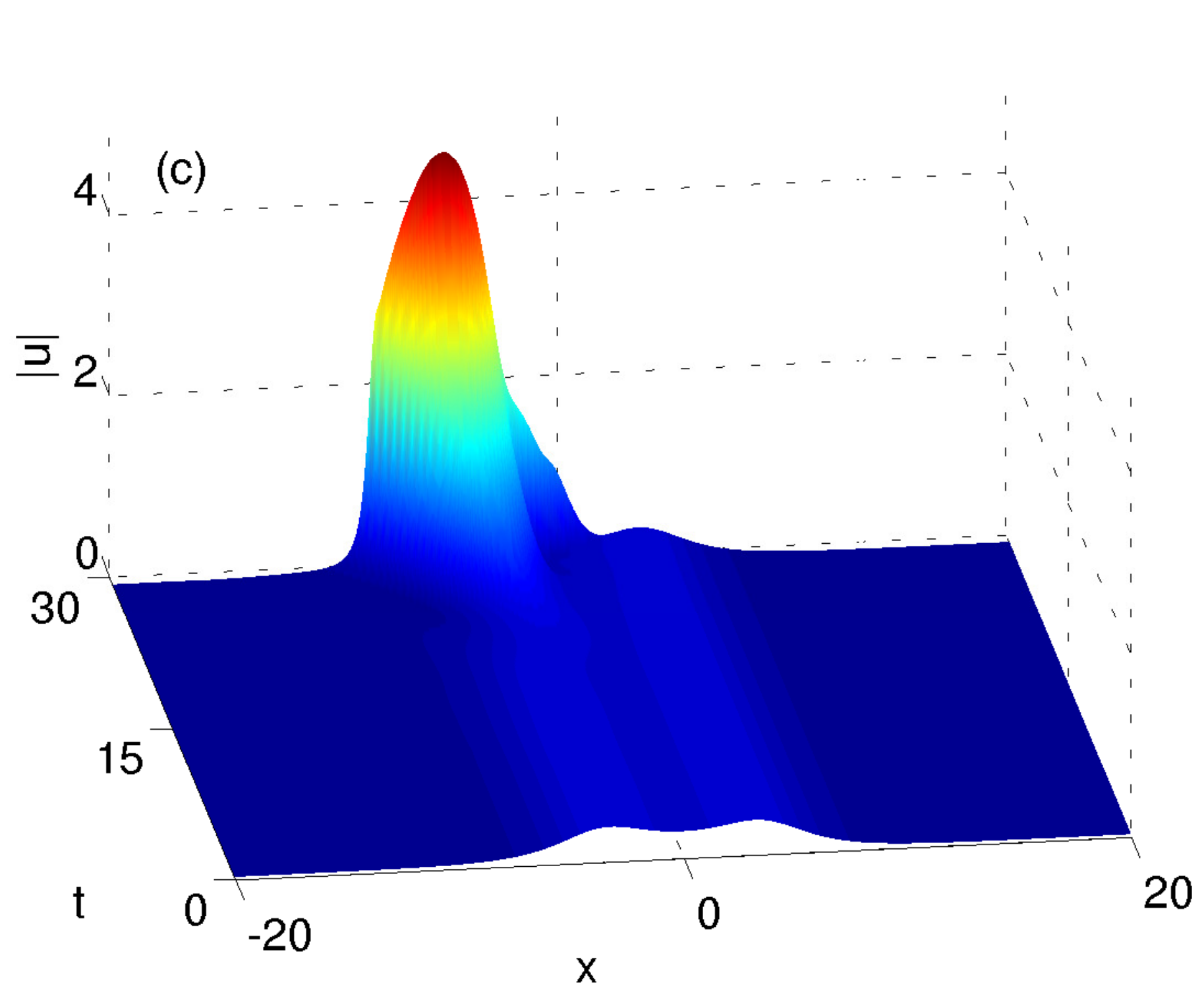} 
  \hspace*{0.02 \linewidth}
   \includegraphics*[width=0.48\linewidth] {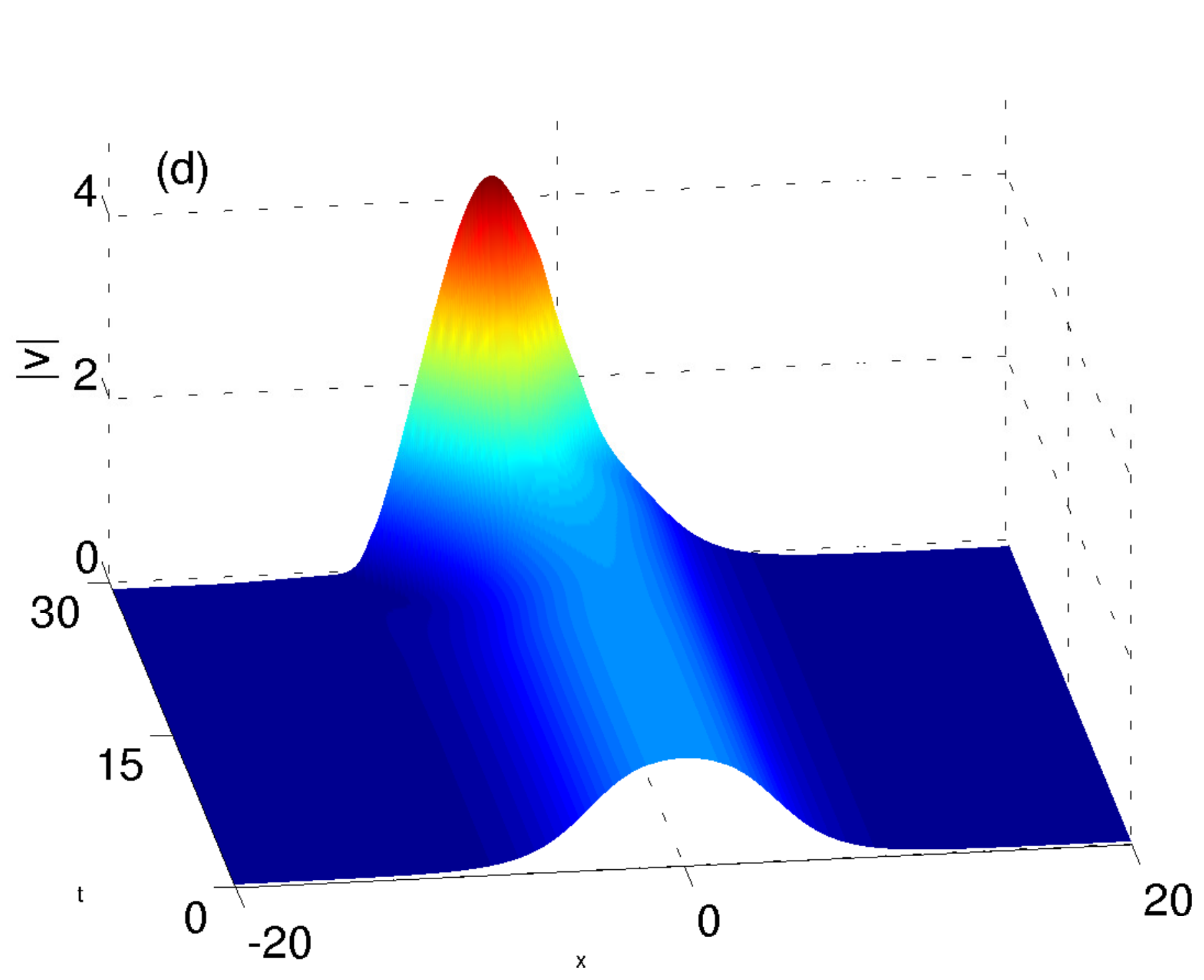} 
  \end{center}
  \caption{\label{instab}
  The   evolution of the soliton with $\omega$  near 
  the boundary $\omega_c(\gamma)$. Panels (a) and (b) correspond to the ($\gamma=0$)-model \eref{A700}. 
  The soliton frequency is
  $\omega= 0.7075$ and the underlying unstable eigenvalues are $ \lambda= 0.019 \pm 1.7 i$.
  The instability results in a seesaw-like rising and falling of the soliton's two humps.
  Panels (c) and (d) pertain to the \PT-symmetric model \eref{F1} with
   $\gamma=0.5$. Here  $\omega=0.653$ and  
  $\lambda=0.33 \pm 0.33i$.
  The soliton's amplitude grows without bound.
   }
 \end{figure}

The conclusions of the spectral analysis were found to be  consistent with the 
 direct numerical simulations of the equations \eref{A700} and \eref{F1}.  In either case
the initial condition was chosen in the form of the corresponding soliton  solution.
Figure \ref{instab} illustrates the growth of the instability of the soliton with $\omega$ near the 
boundary of its existence domain, with $\gamma=0$ and $\gamma \neq 0$.
The simulations were performed using 
Lakoba's method of characteristics  \cite{Taras}.

\section{Concluding remarks} 
\label{Conclusions}

The key results of our paper can be summarised as follows.

(1)
We have formulated 
a recipe for the   \PT-symmetric extension of spinor models that is consistent with the relativistic invariance and
gain-loss interpretation of the \p- and \T-breaking terms.

(2) 
The  \PT-symmetric extension of the massive Thirring model
was shown to be gauge equivalent to the original Thirring model.
This proves that  the \PT-symmetric model is a completely integrable system and has infinitely many conserved quantities. 
We have derived, explicitly,  the first three of these and produced an explicit expression for the soliton solution.

(3) 
We have established the local momentum conservation law for the \PT-symmetric Gross-Neveu equation 
and obtained an exact explicit 
 soliton solution for that model. 

(4) 
A novel nonlinear Dirac equation  was introduced, along with its \PT-symmetric extension.
We have determined an exact soliton solution of the new model.
In the $\gamma=0$ case this solution is explicit 
while in the $\gamma \neq 0$ sector it is obtained as a quadrature. 
The soliton was found to be stable in most of  its  domain  of existence in the $(\gamma, \omega)$ parameter plane.
The instability is only present in a narrow strip along the boundary of this domain.

The central message of this study is that of a remarkable ubiquity 
as well as structural and dynamical stability  of  spinor
solitons. Solitons are supported by the Lorentz-invariant Dirac equations with a broad range of cubic nonlinearities.
They
 persist under the addition of  the \p- and \T-breaking \PT-symmetric terms.
No matter how orderly or disorderly the \PT-symmetric extension is --- whether it has infinitely many conservation laws or none ---
the solitons are expressible in exact analytic form.
 Finally,  the spinor solitons are stable in the 
Lyapunov sense --- either  in the entirety or in the vast majority of their parameter domain,
 both in the original model and in its \PT-symmetric extension.

We close this section with two remarks.

The first one concerns travelling solitons.
In this paper, we have restricted ourselves to considering the relativistically invariant nonlinear Dirac equations. 
Accordingly, each of our stationary localised solutions represents a one-parameter family of travelling solitons which 
can be retrieved 
 by the Lorentz boost \eref{A2}-\eref{A3}. The stability properties of the moving solitons do not depend on their velocity.
 In contrast,  obtaining travelling solitons in the Dirac 
equations with non-invariant nonlinearities would be a nontrivial affair.
Stability would also have to be examined for each value of the velocity individually.

The second remark is on the interpretation of the 
 $\mathcal{P}$- and $\mathcal{T}$-breaking terms.
  In the Schr\"odinger equations governing the amplitudes of optical beams, these terms 
  are associated with gain and loss.
  The particular type of  the \PT-symmetric  Dirac perturbations  that we scrutinised in this study, have a similar nature.
  However the Dirac modes that gain and lose energy, are not the components of the spinor but their linear superpositions.
  (See equation \eref{A640}). 
  In the underlying physical system
  (e.g. two coupled oscillator chains), the symmetry-breaking terms may arise due to the coupling asymmetry rather than 
  plain gain and loss. See section \ref{Diatomic} above.

\section*
{Acknowledgments}
The authors gratefully acknowledge useful discussions with Georgy Alfimov,
 Abdul Kara and Boris Malomed. We thank Dmitry Pelinovsky and  Taras Lakoba for reading the paper and giving their comments.
This work was supported in part by the US Department
of Energy.
NA and IB would like to thank Center for
Nonlinear Studies, Los Alamos National Laboratory, for
warm hospitality during their stay.
NA and IB were  also supported by the National Research
Foundation of South Africa (grants 105835, 85751 and 466082)
and
the European Union's Horizon 2020 research
and innovation programme under the Marie Sk{\l}odowska-Curie
Grant Agreement No. 691011. Computations were performed at the UCT
HPC Cluster.


\newpage 
\begin{thebibliography}{99}


\bibitem{Bender}
 C.M. Bender and S. Boettcher, Phys. Rev. Lett. {\bf 80}   5243 (1998);
 C M Bender, S Boettcher, and P N Meisinger,
  Journ Math Phys {\bf 40} 2201 (1999); 
  Bender C M, 
Contemp. Phys. {\bf 46}  277 (2005);  C M Bender, Rep Prog Phys {\bf 70} 947 (2007)

\bibitem{PT_Focus}
Focus on Parity-Time Symmetry in Optics and Photonics. Editors: 
D Christodoulides, R El-Ganainy, U Peschel, S Rotter. New J Phys (2015-2017);
Issue on Parity Time Photonics.
Editors: V Kovanis,  J Dionne, D Christodoulides,  A Desyatnikov.
IEEE Journal of Selected Topics in Quantum Electronics {\bf 22}, issue 5 (2016)

\bibitem{refraction}
K.G. Makris,  R. El-Ganainy, D.N. Christodoulides, Z.H. Musslimani, 
Phys. Rev. Lett. {\bf 100} 103904 (2008);
 M.C. Zheng,
 D.N. Christodoulides, R. Fleischmann, T. Kottos, 
Phys. Rev. A {\bf 82} 010103 (2010)

\bibitem{Regensburger} 
 A. Regensburger, C. Bersch,  M.-A. Miri, G. Onishchukov,
D. N. Christodoulides, and U. Peschel,
Nature (London) {\bf 488} 167 (2012).


\bibitem{Guo} 
A. Guo, G. J. Salamo, D. Duchesne, R.Morandotti, M. Volatier-Ravat, V. Aimez, G. A. Siviloglou, and D. N. Christodoulides,
Phys. Rev. Lett. {\bf 103} 093902 (2009);
O.V. Shramkova and  G.P. Tsironis,  Scientific Reports {\bf 7}  42919 (2017)



\bibitem{Nonreciprocal_propagation}
O. Bendix, R. Fleischmann, T. Kottos, and B. Shapiro,
Phys. Rev. Lett. {\bf 103}  030402 (2009);
H. Ramezani, T. Kottos, R. El-Ganainy, and
D. Christodoulides, Phys. Rev. A {\bf 82}  043803 (2010);
Peng B, \"Ozdemir \c{S}K, Lei F, Monifi F, Gianfreda M, 
Long G, Fan S, Nori F, Bender CM, Yang L,  Nat. Phys. {\bf 10}, 394 (2014)






















\bibitem{switch}
 M. Kulishov,
J. M. Laniel, N. BŽlanger, J. Aza–a, and D. V. Plant,
Opt. Express {\bf 13}  3068 (2005);
 A.A. Sukhorukov,
 Z.Y. Xu, Yu.S. Kivshar, 
Phys. Rev. A {\bf 82} 043818 (2010);
Z. Lin, H. Ramezani, T. Eichelkraut, T. Kottos, H. Cao, and D. Christodoulides,
Phys. Rev. Lett. {\bf 106}  213901 (2011)







\bibitem{RKEC} H. Ramezani,
 T. Kottos,  R. El-Ganainy, D.N. Christodoulides, 
Phys. Rev. A {\bf 82} 043803 (2010)








\bibitem{invisibility} 
L Feng, Y-L Xu,  W. S. Fegadolli, M-H Lu, J E. B. Oliveira, V R. Almeida, Y-F Chen, and  A Scherer.
 Nature Mater. {\bf 12} 108 (2012);
 L. L. S\'anchez-Soto and J. J. Monzon, Symmetry  {\bf 6} 396
(2014)

\bibitem{plasmonics} H. Benisty,   A. Degiron, A. Lupu, A. De Lustrac,  S. ChŽnais,  S. Forget, M.  Besbes, G. Barbillon, A. Bruyant,  S. Blaize, and G. LŽrondel,  Opt. Express {\bf 19}, 18004 (2011);  M. Mattheakis, Th. Oikonomou, M. I. Molina and G. P. Tsironis,  IEEE J. Sel. Top. in Quantum Electronics {\bf 22} 5000206 (2015)


  \bibitem{OM}  H. Jing,     \c{S}. K.  \"Ozdemir, Z. Geng,  J. Zhang,  X.-Y. L\"u,      B. Peng,  L. Yang, and F. Nori,  Sci. Rep. {\bf 5}  9663  (2015);
  K V Kepesidis,  T J Milburn, J Huber,  K G Makris, S Rotter, and  P Rabl,  New J. Phys. {\bf 18}  095003 (2016) 


\bibitem{Lazarides}  N. Lazarides and G. P. Tsironis,  Phys. Rev. Lett. {\bf 110}   053901 (2013).

\bibitem{NLS_single}  
K.G. Makris,  R. El-Ganainy, D.N. Christodoulides, Z.H. Musslimani, 
Phys. Rev. Lett. {\bf 100} 103904 (2008);
Cartarius H and   Wunner G  2012
{\it Phys. Rev. A} {\bf 86} 013612; 
Cartarius H, Haag D, Dast D and  Wunner D 2012
\textit{J. Phys. A: Math. Theor.} {\bf 45}  444008;  
 D Dast,   D  Haag,    H Cartarius,   G  Wunner,   R  Eichler,  
and J  Main,  Fortschr. Phys. {\bf 61}  124  (2013);
  V V Konotop and D A Zezyulin, Optics Lett {\bf 39} (2014)  5535;
  J Yang, Phys Lett A {\bf 378} (2014) 367; 
  J Yang,   Optics Lett  {\bf 39}  5547 (2014);
    I V Barashenkov,  D A Zezyulin and V V Konotop, New J Phys {\bf 18}  (2016) 075015;
D A Zezyulin, I V Barashenkov and V V Konotop, 
Phys Rev  A {\bf 94} 063649 (2016);
D A Zezyulin, Y V Kartashov and V  V Konotop,  Optics Letters {\bf 42} 1273 (2017)





\bibitem{NLS_dimer}
 S V Suchkov, B A Malomed, S V Dmitriev, Y S Kivshar,  Phys Rev E {\bf 84}, 046609 (2011);
 R. Driben and B.A. Malomed, Opt. Lett. {\bf 36} 4323 (2011); 
  N V Alexeeva, I V Barashenkov, 
A A Sukhorukov, and Y S Kivshar,  Phys. Rev. A {\bf 85} 063837 (2012);
I.V. Barashenkov, 
S. V. Suchkov,  A. A. Sukhorukov, S. V. Dmitriev, and Yu. S. Kivshar,
Phys. Rev. A {\bf 86} 053809 (2012)


\bibitem{NLS_disc}
V. V. Konotop, D. E. Pelinovsky,  and D. A. Zezyulin, EPL {\bf 100} 56006  (2012);
I.V. Barashenkov, L. Baker, and N.V. Alexeeva, Phys Rev A {\bf 87} 033819 (2013);
D E Pelinovsky, D A Zezyulin, V V Konotop, Journ Phys A: Math Theor {\bf 47} 085204 (2014);
A Chernyavsky and D E Pelinovsky, J. Phys. A: Math. Theor. {\bf 49}      475201      (2016);  A Chernyavsky and D E Pelinovsky, 
Symmetry  {\bf 8}  59  (2016);
N V Alexeeva, I V Barashenkov, and Y S Kivshar.
New J Phys {\bf 19}  (2017) 113032



\bibitem{reviews} 
V V Konotop, J Yang, and D A Zezyulin, Rev. Mod. Phys. {\bf 88} 035002 (2016);
S V Suchkov, A A Sukhorukov, J H Huang, S V Dmitriev, C Lee, and Y S Kivshar,
Laser and Photonics Reviews {\bf 10} 177 (2016) 


\bibitem{R1}  D. Ivanenko, ZhETF {\bf 8}  260 (1938)
\bibitem{R3}   W.E. Thirring, Ann. Phys. {\bf 3} 91 (1958)
\bibitem{R6} D D Ivanenko and M M Mirianashvili,  DAN SSSR {\bf 106} 413 (1956)
 \bibitem{R2} W. Heisenberg, Rev. Mod. Phys. {\bf 29} 269 (1957)

\bibitem{R4}        M. Soler, Phys. Rev. D {\bf 1} 2766 (1970)
\bibitem{R5}       D.J. Gross and A. Neveu, Phys. Rev. D  {\bf 10} 3235 (1974)  

\bibitem{BC} D K Campbell and A R Bishop, Phys Rev B  {\bf 24} 4859 (1981);
Nucl Phys B {\bf 200} 297 (1982)




\bibitem{gratings}  A B Aceves and S Wabnitz, Phys Lett A {\bf 141} 37 (1989); 
C. M. de Sterke and J. E. Sipe, in Progress in Optics,
edited by E. Wolf (Elsevier, Amsterdam, 1994), Vol.
XXXIII; C. M. de Sterke, D. G. Salinas, J. E. Sipe, Phys Rev E {\bf 54}  1969 (1996)




\bibitem{Mikhailov}
A. V. Mikhailov, JETP Lett {\bf 23} 320 (1976);
E. A. Kuznetsov   and A. V. Mikhailov,  Theor. Math. Phys. {\bf 30} 303 (1977);
D J Kaup and  A C Newell, Lett Nuovo Cim {\bf 20} 325 (1977)

\bibitem{R8}       B. Feng, O Sugino, R-Y Liu, J Zhang, R Yukawa, M Kawamura, T Iimori, H Kim, Y Hasegawa, H Li, L Chen, K Wu, H Kumigashira, F Komori, T-C Chiang, S Meng, and I Matsuda,
 Phys. Rev. Lett. 118, 096401 (2017)
 
\bibitem{R9}      K F  Mak, C Lee, J Hone, J Shan, and T F Heinz,
 Phys. Rev. Lett. 105, 136805 (2010)

\bibitem{R7}       L.H. Haddad and L.D. Carr, New J. Phys. 17, 113011 (2015).  


\bibitem{R10}      M.J. Ablowitz and Y. Zhu, Phys. Rev. A 82, 013840 (2010). 
\bibitem{R11}      O Peleg, G Bartal, B Freedman, O Manela, M Segev, and D N Christodoulides,
 Phys. Rev. Lett. 98, 103901 (2007)  
\bibitem{R12}        J Dalibard, F Gerbier, G  Juzeli\={u}nas, and P  \"Ohberg,
Rev. Mod. Phys. 83, 1523 (2011). 




\bibitem{new_stability}
M Chugunova and D  Pelinovsky, SIAM J. Applied Dynamical Systems {\bf 5} 66 (2006);
F Cooper, A Khare, B Mihaila and A Saxena, Phys Rev E {\bf 82} 036604 (2010);
 N. Boussaid  and A. Comech,       	arXiv:1211.3336 [math.AP] 2012;
 N. Boussaïd and  S. Cuccagna, Commun. Partial Differ. Equ. {\bf 37}  1001 (2012);
 A Comech,  M Guan,   S  Gustafson, Annales de l'Institut Henri Poincare (C) Non Linear Analysis
 {\bf 31} 639 (2014); 
 S Shao, N R Quintero,   F G Mertens, F Cooper, A Khare, and A Saxena, Phys Rev E {\bf 90} 032915 (2014);
 G Berkolaiko,   A Comech and A Sukhtayev, Nonlinearity {\bf 28} 577 (2015); 
 D Pelinovsky and Y Shimabukuro, J Nonlinear Sci {\bf 26}  365 (2016);
   N. Boussaid  and A. Comech,   arXiv:1705.05481 [math.AP] 2017
  	
	
\bibitem{Berkolaiko}  	 G. Berkolaiko and A. Comech,    Math. Model. Nat. Phenom. {\bf 7}  13   (2012)
	
\bibitem{Num_GN}
F G Mertens,
 N R Quintero, F Cooper, A Khare, and A Saxena, Phys Rev E {\bf 86} 046602 (2012);
J. Cuevas-Maraver, P.G. Kevrekidis, A. Saxena, F. Cooper and F.
Mertens,  in Ordinary and Partial
Differential Equations, New York, USA: Nova Publishers, 2015
	


\bibitem{Dima}
D.E. Pelinovsky and Y. Shimabukuro,  Lett. Math.  Phys. {\bf 104}  21 (2014);
A. Contreras, D.E. Pelinovsky, and Y. Shimabukuro,   Commun.  Partial Diff. Equations {\bf 41} 227 (2016)

	
\bibitem{Taras} 
T I Lakoba, Phys Lett A {\bf 308} 300 (2018) 



\bibitem{old_stability} I L Bogolubsky, Phys Lett A {\bf 73} 87 (1979);
A Alvarez and B Carreras, Phys Lett A {\bf 86} 327 (1981);
J Werle, Acta Phys Polonica B {\bf 12} 601 (1981);
A Alvarez and M Soler, Phys Rev Lett {\bf 50} 1230 (1983);
P Mathieu and T F Morris, Phys Lett B {\bf 126} 74 (1983);
P Mathieu and T F Morris, Phys Lett B {\bf 155} 156 (1985);
W A Strauss and L V\'azquez, Phys Rev D {\bf 34} 641 (1986)



\bibitem{BPZ} 
I V Barashenkov, D E Pelinovsky and E V Zemlyanaya, Phys Rev Lett  {\bf 80} 5117 (1998) 


\bibitem{Akhmediev}
N Akhmediev and A Ankiewicz, Phys Rev Lett {\bf 70} 2395 (1993);
J   M Soto-Crespo and N Akhmediev,       Phys Rev E {\bf 48} 4710    (1993);
N Akhmediev and  J M Soto-Crespo, Phys Rev E {\bf 49} 4519 (1994);
V  Rastogi,  K S  Chiang,  N N  Akhmediev,  Phys Lett A {\bf 301}      27 (2002) 


\bibitem{Bender_spinor} 
C  M  Bender,  H F  Jones,  R J  Rivers,
Phys Lett B {\bf 625}  (2005) 333;



\bibitem{Cuevas}
J. Cuevas-Maraver, P.G. Kevrekidis, A. Saxena, F. Cooper, A. Khare, A. Comech, and C. M. Bender,
IEEE: J. Selected Topics in Quantum Electronics {\bf 22}  5000109 (2016)


\bibitem{Sakaguchi}
H Sakaguchi and B A Malomed, New J Phys {\bf 18} (2016) 105005


\bibitem{diatomic}
Y S Kivshar, N Flytzanis, Phys Rev A {\bf 46}  7972 (1992); 
Y S Kivshar, O A Chubykalo, O V Usatenko, D V Grinyoff, Int. J. Mod. Phys. B {\bf 9} 2963 (1995)



\bibitem{David}  D David, J. Math. Phys. {\bf 25} 3424 (1984)

\bibitem{BG} I V Barashenkov and B S Getmanov,  Commun. Math. Phys. {\bf 112} 423 (1987)



\bibitem{LKG} 
S. Y. Lee, T. K. Kuo, and A. Gavrielides, Phys Rev D {\bf 12} 2249 (1975)






\bibitem{BPD} I V Barashenkov, D E Pelinovsky and P Dubard,
Journ Phys A: Math Theor {\bf 48} (2015) 325201

\end{thebibliography}
\end{document}